\documentclass[letterpaper, 10 pt, onecolumn, conference]{IEEEtran}  % Comment this line out
\usepackage[none]{hyphenat}

\renewcommand{\baselinestretch}{1.5} %line spacing 1.5

\usepackage[linktoc=all]{hyperref}
\hypersetup{pdftex,colorlinks=true,allcolors=blue}
\usepackage{hypcap}

\usepackage{graphics} % for pdf, bitmapped graphics files
\usepackage{epsfig} % for postscript graphics files

\title{\LARGE \bf
FlexNGIA: A Flexible Internet Architecture \\for the Next-Generation Tactile Internet
}
\author{ \parbox{3.4in}{\centering Mohamed Faten Zhani\\%*
         \'Ecole de Technologie Sup\'erieure (\'ETS Montreal) \\
         Montreal, Quebec, Canada\\
				  }
         \hspace*{ 0.15 in}
         \parbox{3in}{ \centering Hesham ElBakoury\\ %**
         Futurewei Technologies, Inc. \\
         Santa Clara, California, USA\\
				}
}

\begin{document}

\maketitle
\thispagestyle{empty}
\pagestyle{empty}

\providecommand{\keywords}[1]
{
  \small	
  \textbf{\textit{Keywords---}} #1
}

%%%%%%%%%%%%%%%%%%%%%%%%%%%%%%%%%%%%%%
%%%%%%%%%%%%%%%%%%%%%%%%%%%%%%%%%%%%%%

%The~keynote details are available at \url{https://goo.gl/ppdsVP}.}

\begin{abstract}
From virtual reality and telepresence, to augmented reality, holoportation, and remotely controlled robotics, these future~network applications promise an unprecedented development for society, economics and culture by revolutionizing the way we live, learn, work and play. In order to deploy such futuristic applications and to cater to~their~performance requirements, recent trends stressed the need for the ``Tactile Internet'', an Internet that, according to~the~International Telecommunication Union (ITU), combines ultra low latency with extremely high availability, reliability and security~\cite{ITUTactileInternet}. Unfortunately, today’s Internet falls short when it comes to~providing such stringent requirements due to several fundamental limitations in the design of~the~current network architecture and~communication protocols. This brings the need to rethink the network architecture and~protocols, and~efficiently harness recent technological advances in terms of virtualization and network softwarization to design the Tactile Internet~of~the~future.
 
In this paper, we start by analyzing the characteristics and requirements of future networking applications. We~then~highlight the limitations of the traditional network architecture and protocols and their inability to cater to~these~requirements. Afterward, we~put~forward a~novel network architecture adapted to the Tactile Internet  called FlexNGIA, a Flexible Next-Generation Internet Architecture\footnote{Details are available on \url{https://www.FlexNGIA.net} \\FlexNGIA was presented at~the~keynote of ACM SIGCOMM 2019 Workshop on Networking for Emerging Applications and Technologies (NEAT/SIGCOMM 2019) (\url{http://bit.ly/31W6vb2})
and at the keynote of~the~IEEE/IFIP/ACM~International Workshop on~High-Precision Networks, Operations and Control (HiPNet/CNSM 2018), Rome, Italy, November 5-9, 2018 (\url{https://goo.gl/ppdsVP}}. 
%The keynote recording is available on YouTube at \url{https://youtu.be/k5MCtxxzbKk
We then describe some use-cases where we discuss the potential mechanisms and control loops that could be offered by FlexNGIA in order to ensure the required performance and reliability guarantees for future applications. Finally,~we~identify the key research challenges to further develop FlexNGIA towards a full-fledged architecture for the future Tactile Internet. \\\\


% TCP \and  QUIC \and  SCTP \and  FlexNGIA \and  Holoportation \and  Telepresence \and  Virtual Reality \and  Augmented Reality}
\end{abstract}

%Table of content
\newpage
\tableofcontents
\vspace{1.5cm} 
%\newpage

\addtocontents{toc}{\protect\setcounter{tocdepth}{2}} %hide subsubsection

%------------------------------------------------------
\section{Introduction}
%------------------------------------------------------
The last few years have witnessed the emergence of a new breed of futuristic applications. Ranging from virtual reality, to~augmented reality, holoportation, and~telepresence, these applications are bound to completely revolutionize different domains including communications, healthcare, education, commerce, gaming and culture. 
Unfortunately, the~deployment of these applications has been significantly inhibited because of~the~inability of~today's Internet technologies and protocols to cater to~their~requirements in terms of performance (e.g.,~latency, bandwidth), reliability, and~availability.
% and to adapt the network services to the characteristics where the performance and availability of the communications services are not guaranteed
%especially at large scale networks . This is mainly due to
%
%Unfortunately, the inability of today's Internet technologies and protocols to cater to the performance requirements of these applications has been a significant inhibitor to their expansion and deployment, especially at large scale networks where the performance and availability of the communications services are not guaranteed. 
Indeed, the~current Internet architecture and communication protocols suffer from several fundamental limitations in~their~design, which does not allow them to offer the requirements of these future applications, which leads to low performance (e.g.,~high latency and~packet loss) and poor user experience.
%we donèt trust internet for performance, availability, reliability.
%
%Unfortunately, deploying and running such applications over today's Internet lead to decreased performance and poor user experience, due to several fundamental limitations in the design of~the~current network architecture and communication protocols. 
%In addition to today's limitations, the emerging of new technologies and paradigms like virtualization and~Software Defined Networking~(SDN), is paving the way to
%Propelled by the need to overcome these limitations, the emerging of new technologies and paradigms like virtualization and~Software Defined Networking~(SDN)

Propelled by the need to overcome these limitations and motivated by the emergence and maturity of new technologies and~paradigms like virtualization, ~software defined networking, and network programmability, there is a growing determination in the research and industrial communities to rethink the architecture of the Internet and  to design novel communication protocols and network services. The ultimate goal is~to~build the next-generation ``Tactile'' Internet, that would offer an ultra-low latency with high levels of performance, availability, resiliency and~security \cite{ITUTactileInternet}.

In this paper, we propose a novel fully-\textbf{Flex}ible \textbf{N}ext-\textbf{G}eneration \textbf{I}nternet \textbf{A}rchitecture (called FlexNGIA) where we provide our~vision of the future Tactile Internet infrastructure and services, business model, management framework, and network protocol stack. More specifically, the FlexNGIA architecture is characterized by~the~following features: \\
\textbf{$\bullet$ Business and Service Model:} FlexNGIA defines the different stakeholders that should be involved in the future Internet and~identifies the~services that should be offered by future network operators (i.e.,~Internet service providers).
We argue that future networks have to offer not only data delivery service but rather Service Functions Chains (SFCs) that are able to~carry the~traffic between multiple sources of data to~multiple destinations and~to~offer network functions that~are~tailored to~the~application needs. These chains should also be customizable and able to~offer different level of~performance, availability and~reliability guarantees that could be customized based on the applications' requirements.\\
\textbf{$\bullet$ In-network computing:} FlexNGIA promotes the deployment of computational resources throughout the infrastructure from the edge to the edge including the core of the network. This provides the network with a real full programmability and~an~unprecedented flexibility for operators to deploy a wide array of network functions within the network. \\
%which should not be restricted to the programmability in~terms of~the~forwarding rules (as it is currently the case for current SDN OpenFlow-based switches) but~also the programmability of the services offered by the network functions that could operate at different layers of~the~network stack (e.g.,~application, transport or network layers). \\
%. As a result, the network could play a more crucial role in supporting network functions at the different layers of 
%
\textbf{$\bullet$ Basic and advanced network functions:} thanks to in-network computing, FlexNGIA considers a wide range of network functions including the basic functions (e.g.,~packet forwarding, routing, firewall) but also more advanced functions to~support the~applications (e.g.,~data compression, video and hologram processing) and the transport layer (e.g.,~congestion control, video cropping). This~requires to~broaden the concept of Software-Defined Networking to provide not only APIs and protocols to~define basic forwarding rules (e.g.,~OpenFlow~\cite{McKeown_OpenFlow}) or data plane programmability (e.g.,~P4~\cite{BosshartP42014}) but also APIs and protocols to control the operation of the aforementioned advanced network functions.\\
\textbf{$\bullet$ Cross-layer network protocol design:} the FlexNGIA architecture advocates to combine the transport and network layers in~order to~ensure a~better control  over the services of layers 3 and 4 and further improve their~efficiency. The  FlexNGIA combined layer breaks the end-to-end principle adopted by today's Internet. However, this allows the network to offer services like data reliability (e.g.,~packet loss detection and retransmission) and~congestion control in~order~to~improve the~overall performance in terms of latency, bandwidth and packet loss. \\
\textbf{$\bullet$ Application-aware networking:} the FlexNGIA design allows the network to be aware of the application. In~other~words, the~network (i.e.,~the~combination of~transport and network layers)
 is aware of the flows used by the same application even~if~the~application's components are distributed and~sending~data from~different sources to~different~destinations. 
The~network is~then~aware of the run-time changing requirements of each flow in terms of performance, reliability and~availability. It~is~also aware of~the~type of data being transmitted and could take decisions taking into account the application context and~the~user behavior within the application (e.g.,~cropping a~video to~show only the~important objects of~a~video or~a~hologram).\\
\textbf{$\bullet$ Flexible application design and traffic engineering:} The FlexNGIA architecture allows the application designer to customize the service function chain adapted to each of the deployed applications and to define the appropriate routing strategy within this chain. For instance, routing could be run in a centralized manner like in software-defined networks or could use other strategies like Segment Routing \cite{rfc8402SegmentRouting} where packets contain different kind of instructions (including forwarding instructions). \\
\textbf{$\bullet$ Simplified Layer 2 virtualization:} the FlexNGIA architecture relies on a layer-2 virtualization that allows to easily separate the traffic of the different network applications. This can be seen as an extension of the VLAN technology that would allow to identify each application of the network with a unique ID that remains valid even when packets travel through several networks managed by different network operators.\\
\textbf{$\bullet$ Fully flexible packet header format tailored to each application:} FlexNGIA promotes a totally flexible header for~the~upper layers (i.e.,~layer~3 and~above) that could be defined by the application designer depending on the application type and~requirements. The~headers could include meta-data (i.e.,~different kind of information) and also commands to be used or~executed by~the~network functions or the network resource management framework.

In the remaining part of this paper, we start by describing some of the major future Internet applications and we analyze their~key characteristics and requirements (Section~\ref{section:GlanceIntoFutureInternetApplications}). 
We~then~present and dissect the limitations of the today's network architecture, protocol stack, packet headers and~sources of latency, and we show how they are not able to cater to these requirements (Section~\ref{sec:COfTradNet}). We next draw a~rough sketch of~the~proposed network architecture adapted to the Tactile Internet (FlexNGIA) where we provide our vision of~the~next-generation network infrastructure and services, business model, management framework, network protocol stack and headers (Section~\ref{sec:FlexNGIA}). 
We describe afterward some use-cases where we discuss the~potential mechanisms and control loops that are offered by~FlexNGIA in order to ensure the required performance and~reliability guarantees for future applications (Section~\ref{sec:UseCases}). 
We~then summarize the key research challenges that should be addressed to~further develop FlexNGIA towards a full-fledged architecture for the Tactile Internet (Section~\ref{sec:KeyResearchChallenges}).
We finally wrap~up with~some concluding remarks to highlight the main features of FlexNGIA (Section~\ref{sec:Conclusion}).

%In this document, we highlight the limitations of traditional network architectures to satisfy the performance requirements of such applications and then present the research opportunities and the key challenges to be solved in order to address such limitations. We finally present sketches of some potential solutions that should be further developed to design next-generation network architectures and infrastructure management framework able to cope with these applications' requirements.

%------------------------------------------------------
\section{A Glance into Future Internet applications}\label{section:GlanceIntoFutureInternetApplications}
%------------------------------------------------------
In the last few years, we are witnessing the emergence of a new breed of futuristic applications that are gaining momentum. From virtual reality, holoportation, augmented reality, to telepresence, and remotely-controlled robotics, these applications are changing the IT landscape and are pushing towards a new generation of the Internet that should be able to accommodate the~high-performance and changing requirements of these applications. In the following, we provide examples of~typical futuristic applications and~then~analyze their characteristics and requirements.

%-------------------------------------------------------------------------
\subsection{Typical Future Applications} \label{sec:TypicalAppsGlanceIntoTheFuture}
%-------------------------------------------------------------------------
Before delving into the technical issues pertaining to next-generation networks, we start in the following by~describing some~potential~network applications that will be~common in the future: 

$\bullet$ \textbf{Telepresence:} this application allows to navigate through streets, buildings, meet people, or carry out high-precision tasks via a telepresence robot. This assumes that the user is able to remotely control the robot in a perfect and precise manner and has a~3D~high-definition visual display of the robot environment, a perfect 3D surrounding sound and~ideally kinesthetic communications with Haptic technologies that recreate the sense of touch perceived by the robot (by~applying pressure, vibrations, or motions to the user). Telesurgery, which refers to the ability for a doctor to perform a~surgery remotely using a robot, can be seen as a form of telepresence. This is~a typical~application that requires high guarantees of performance (e.g.,~in~terms~of~bandwidth and~latency) and~reliability.

$\bullet$ \textbf{Virtual Reality:} this application allows the user to navigate through a virtual environment and interact with virtual objects. Similar to telepresence, it requires to recreate this environment including 3D visual display, sound, and haptic senses.

$\bullet$ \textbf{Holoportation:} holograms refers to high quality 3D models of humans or objects.
Holoportation is defined as~the~technology that allows holograms to be captured, compressed then transmitted and reconstructed at real-time in a distant location. The~application~should allow the ported hologram to be incorporated into a virtual or real environment and interact with~objects and humans located in this environment.

$\bullet$ \textbf{Augmented reality:} this is a technology that allows to superimpose a computer-generated objects on a user's view of~the~real world. These~objects could be images, videos, holograms, text or results of some analysis carried out on the current physical environment of~the~user. Of course, this brings the challenge of capturing the real-world environment of the user, analyzing it and providing the outcome of the analysis in real-time.

In a near future, we can expect to have these applications combined and to witness the merge of physical and digital worlds where virtual and real become part of our environments. It is also expected that these future applications incorporate haptic, taste and smell communications in addition to vision and audition in~order~to~allow an~immersive experience involving the~five senses~\cite{Obrist:2016}. In the following, we provide an example of a killer app that could combine all these features. This application will~be used as a reference application throughout this paper. 

$\bullet$ \textbf{The killer app}: A typical application in the future would be a ``virtual coffee shop" where users can meet each other in a~virtual reality environment that~incorporates their holograms. Users should have an~immersive experience where they can ``live" within this virtual reality environment (i.e.,~the~coffee shop) and interact with each other through their respective holograms as if they were in a real coffee shop. Ideally, they should be able to perfectly use and leverage the human five senses (Sight, Sound, Smell, Taste, and Touch) to interact with each other and with the environment.

Of course, all the aforementioned applications are provided as examples but the future definitely hides much more exciting applications that are still unforeseen at the current time. 

%-------------------------------------------------------------------------
\subsection{Requirements of Future Applications} \label{sec:RequirementsAGlanceIntoTheFuture}
%-------------------------------------------------------------------------

As a matter of fact, the aforementioned applications have performance and reliability requirements that can vary from one application to another. In the following, we briefly summarize these requirements:

$\bullet$ \textbf{Bandwidth:} these applications need to transmit tremendous amounts of data that can go from few tens of megabytes to~terabytes of data per second. Hence, it is clear that there is a pressing need for high bandwidth infrastructures. Table~\ref{TableAppBWConsumption}~provides the amount of bandwidth requirements for a real-time transmission of a single hologram and the requirement for 4K, 8K or~16K virtual environment (16K is the~resolution of~the~human eye retina that could offer a~truly immersive experience for virtual reality~\cite{16KResolution}). It can be seen in the table that a single hologram needs up to 50Mbps even with compression. 
In~more~realistic use-cases with multiple holograms and complex virtual environments (with hundreds of~objects), the~needed amount of bandwidth will increase exponentially to~hundreds of~megabits. 
%-----------------------------------------------
\begin{table}[bthp]
\centering
\caption{Bandwidth requirements for some future applications}
\label{TableAppBWConsumption}
\begin{tabular}{|l|l|}
\hline
\textbf{Application}      & \textbf{Bandwidth}                                                                                             \\ \hline
Holoportation             & \begin{tabular}[c]{@{}l@{}}100 Gbps to 1 Tbps (raw) \cite{HologramsBW2011}\\ 30 to 50 Mbps (compressed) \cite{MicrosoftHoloportation}\end{tabular} \\ \hline
Virtual/Augmented Reality & \begin{tabular}[c]{@{}l@{}}(4K x 4K, 60 fps) $\rightarrow$ 20+ Mbps \cite{Arris2016}\\ 
(8K x 8K, 120 fps) $\rightarrow$ 85+ Mbps \cite{8KResolutionBW2}\\ 
(16K x 16K, 240 fps) $\rightarrow$ 300+ Mbps \cite{Arris2016}
\end{tabular}               \\ \hline
\end{tabular}
\end{table}

% https://uploadvr.com/abrash-2018-predictions-oc5/: 140-degree field of view (FOV), variable depth of focus, and a 4k x 4k panel resolution with 30 pixels per degree density. 

$\bullet$ \textbf{Processing power:} as mentioned earlier, future applications generate and consume tremendous amounts of data that, depending on the application requirements, need to be processed at run-time before or after their transmission or even on their way to the destination. Naturally, the~type of processing will mainly depend on the application type (e.g.,~object recognition, movement detection, data compression, data mining, video manipulation, video rendering, encryption).

$\bullet$	\textbf{Latency}: future applications are highly interactive and hence require high-precision, predictable and~ultra-low latency. Latency can go from 1ms to~1s~depending on the interactivity of the application and the~participating human sense \cite{Fettweis2014,VegaHipNet2018}. For~instance, according~to~Fettweis~et~al.~\cite{Fettweis2014}, the human sense of touch reacts within a range of 100ms to 1second depending on how prepared is the user (i.e.,~if~the~user is~expecting to touch an object the reaction time is around 100ms);  %to muscular reaction 
The human auditory reaction time is between 70ms and~100ms; The~visual sense requires less than 10ms latency between successive pictures; 
Tactile or haptic actions using human limbs require instantaneous visual and audio feedback with a latency of 1ms. This would be the ultimate use-case where the user is manipulating objects in a virtual environment or through a telepresence robot.

Furthermore, some future applications may require multiple flows to be synchronized (i.e.,~multi-flow synchronization), that~is~to~ensure that the generated data from different sources arrive at the destination~within a~specific interval of~time or~even at~a~particular point of~time. As~this~definition suggests, these flows might originate from different sources, which~makes multi-flow synchronization a~daunting challenge requiring high-precision control of~the~data delivery over~the~network.

$\bullet$	\textbf{Reliability}: reliability refers to the guaranteed delivery of the transmitted data. some future applications may not tolerate packet loss and may require high reliability; This means that data cannot be lost and should necessarily be received by~the~destination. It is also possible that some applications require partial reliability, which means that reliable transfer is needed only for some of the transmitted data or only during some period of time.

$\bullet$	\textbf{Availability}: availability referes to~the~proportion of~time for which the application is~available. Many of the aforementioned future applications require ultra-high availability. Some of today's networks offer an availability close to four or five nines~\cite{GoogleCloudAvailability} (~99.99\%~and~99.999\%~correspond to downtimes of 360ms and 4ms per~hour, respectively). However, four or even five nines are definitely not acceptable for~several applications like tele-surgery and autonomous driving where a 4ms downtime could~be~fatal. Achieving high availability requires high resiliency of the infrastructure and the network services, that is a the ability to recover seamlessly from potential software and~hardware failures.
% https://landing.google.com/sre/sre-book/chapters/availability-table/
%https://landing.google.com/sre/sre-book/chapters/embracing-risk/#risk-management_measuring-service-risk_time-availability-equation

%is some of these applications may not tolerate packet loss and may require guaranteed reliability; This means that some data cannot be lost and should be received by the destination.
%-------------------------------------------------------------------------
\subsection{Characteristics of Future Applications} \label{sec:CharacteristicsAGlanceIntoTheFuture}
%-------------------------------------------------------------------------
Looking at the aforementioned application, we can notice that they share many characteristics that could be summarized as~follows:

$\bullet$	\textbf{Octopus-like applications:} future applications requires to open a large number of traffic flows from multiple sources to multiple destinations. These flows are typically coming from different sensors or objects and are directed towards several destinations. In~addition, each of~these~flows may have different performance requirements in terms of~latency (end to end delay and time synchronization), packet loss and throughput.
A typical example of an octopus-like application would be the aforementioned ``virtual coffee shop" application where a virtual reality environment and the user holograms should be transmitted at~real time to~each and every user. Each user in this application is a source of~a~flow of~data towards the~others in order to transmit his own hologram. At~the~same time, he is the destination of~multiple flows coming from the other users. This~results in~an~octopus-like application (e.g.,~similar to an octopus with multiple arms connecting several end points).

$\bullet$	\textbf{Changing requirements:} for these applications, each traffic flow is probably associated with an object or a sensor. Depending on how the context of~the~application evolves over time, the performance requirements of the flows coming from different objects/sensors might change over time. For instance, at some point of time, the flow may need stringent latency requirement with minimal packet loss. At another point of time, it may tolerate more relaxed requirements, i.e., higher latency and packet loss. 
As an example, recall~the~``virtual coffee shop" application. Assuming a~user~A~is~interacting (e.g.,~talking) with user B within the virtual reality environment. Thus,~the~traffic flows between user~A~to~user~B should have stringent requirement in terms of reliability, bandwidth, latency and packet loss. If, at some point of time, user~A~stops interacting with~user~B, these requirements could be relaxed.% as both users should only see each other in the virtual environment but~they~do~not~interact~directly.

Given the aforementioned requirements and characteristics of future applications, we discuss, in the next Section, the~main limitations of today's Internet and why it could not be the ``Tactile Internet'' that~is~able to~support these new breeds of~applications.

%-----------------------------------------------------------------
\section{Limitations of Today's Internet} \label{sec:COfTradNet}
%--------------------------------------------------------------------
%In this section, we mainly focus on the limitations of the today's internet. We start by...... 
The goal of this Section is to identify the limitations of today's Internet and describe how 
%they make it unable to provide the performance requirements needed for future applications.
they impede its ability to~support future applications.
We hence analyze first today's Internet infrastructure and services, the network stack protocols and~headers (i.e.,~the transport and~the~network protocols) then we identify and~discuss the current sources of latency.
%challenges related to transport and network layers as well as the management of the network infrastructure:

%-------------------------------------------------------------------------------
\subsection{Internet Infrastructure and Services} \label{sec:InternetInfrastructureServices}
%-------------------------------------------------------------------------------
%----------------------------------------------------------------------------
\begin{figure}[htbp]
	\centering
		\includegraphics[width=0.40\textwidth]{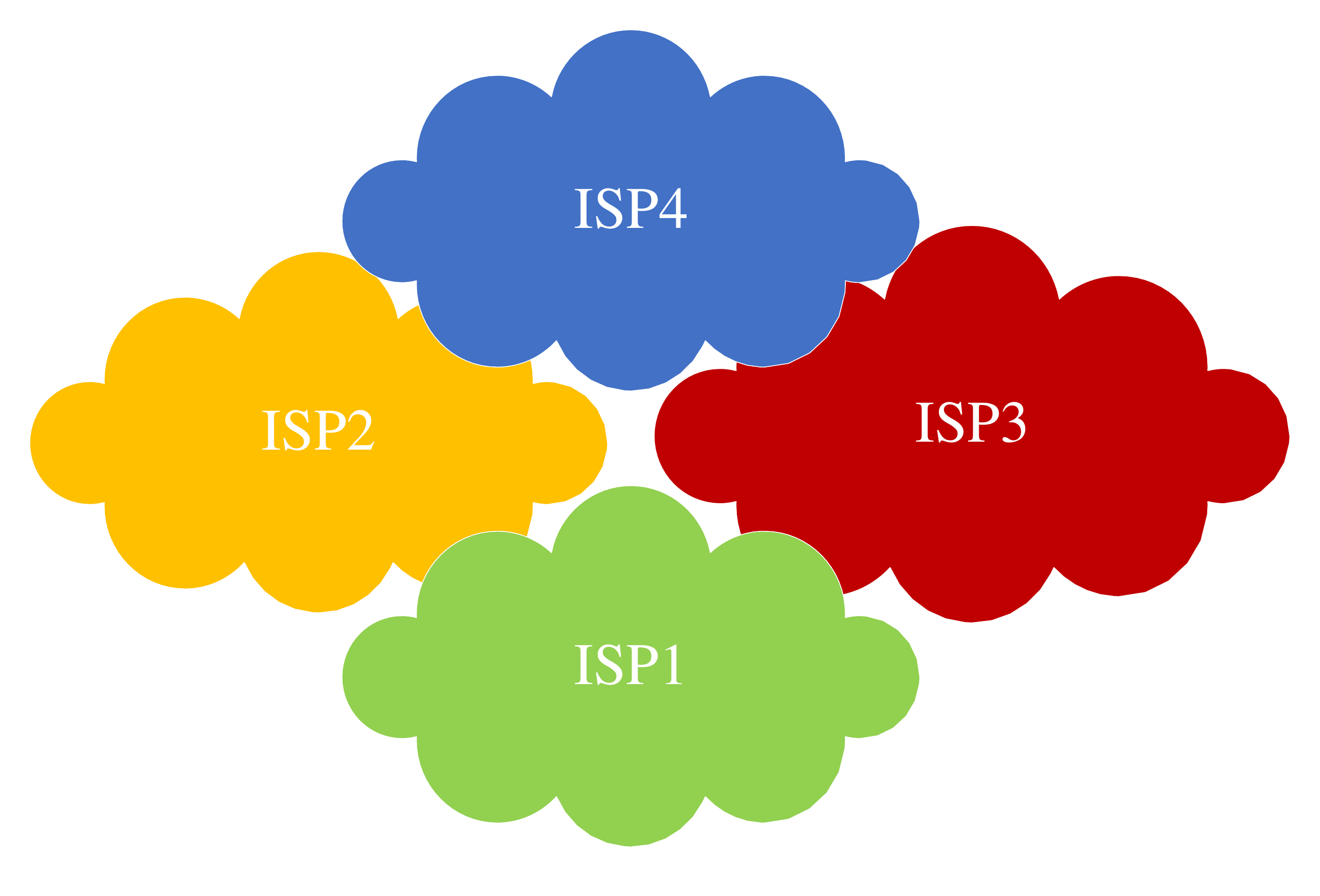}
	\caption{Traditional Internet}
	\label{fig:InternetTraditional}
\end{figure}
%----------------------------------------------------------------------------
%https://en.wikipedia.org/wiki/Internet_transit
%https://en.wikipedia.org/wiki/Internet_service_provider
% https://www.telstraglobal.com/uk/insights/blogs/blog/peering-vs-transit
As shown in Fig.~\ref{fig:InternetTraditional}, today's Internet is a an interconnected system of networks that are operated and managed by~different Internet Service Providers (ISPs). The main and only service provided today by an ISP to the other ones is~the~transit service, that~is~to~allow their traffic to~``transit''~or cross its~network. However, ISPs do not provide guarantees on performance (e.g.,~throughput, delay, packet loss) and availability. 
They also do not offer any function or~service related to~reliability, security, privacy and~application.
It is clear that today's Internet, with its basic ``best-effort'' data delivery service,~cannot be~relied upon to cater to the stringent performance requirements of future applications.

It is also worth noting that Internet today could provide Diffserv \cite{rfc2475DiffServ} which supports different classes of service. However, it does not guarantee deterministic delays for packet delivery and time synchronization of multiple flows.
%-------------------------------------------------------------------------------
\subsection{Transport Layer Protocols} \label{sec:TransportControlManagement}
%-------------------------------------------------------------------------------

Today's Internet transport layer mainly rely on Transport Control Protocol (TCP) \cite{rfc793TCP}  and User Datagram Protocol (UDP)~\cite{rfc768UDP} to ensure end-to-end communications over the network. UDP is a basic protocol that provides only multiplexing and demultiplexing services that allow to identify the source and destination processes at the end hosts. However, TCP offers more services (e.g.~reliability, flow and~congestion control) and~is the most used transport protocol today. That is why, in the following, we~focus on the limitations of TCP (the Reno version). We then compare it to recently proposed transport protocols like SCTP and QUIC and~show how they still inherit many of these limitations. 

$\bullet$ \textbf{One-size-fits-all service offering:} Compared to UDP, TCP provides additional services like reliable data delivery service (i.e.,~error check and retransmission in case of packet loss or errors) and congestion and flow control services operated from the end-points. %ZMF SCTP, QUIC
However, if TCP is used for a connection, all these services are mandatory for all the transmitted data and throughout the~duration of~the~whole communication. For future applications, the transfer requirements in terms of reliability, throughput, packet loss and delay of the same flow may vary significantly over time. For instance, the reliability offered by the transport layer may not be mandatory during the whole communication but only during some periods of time depending on the current application's requirements and~context. In this case, packet loss detection and retransmission for a particular flow could be used only for~some packets or only when necessary during the communication. 

Similarly, other TCP services like congestion and flow control might not be needed for some flows or may be needed during a limited interval of time during the communication. For instance, we can imagine a flow that requires a high and constant traffic rate for some period of time but TCP slow start and congestion avoidance schemes are controlling and limiting the data sending rate, which may significantly hurt the performance of the application.

%----------------------------------------------------------------
$\bullet$ \textbf{The two end points limitation:} 
The transport layer, as defined in the OSI model \cite{OSI}, assumes that a single connection involves only two end-points. Existing protocols like TCP and UDP are building on this definition and provide all their services based on this assumption. As~a~result, an application that is using and manipulating different objects could either use a single TCP connection to transfer the data associated with all these objects or use multiple TCP connections to transport these data. In~addition, when these objects are not in the same physical location, there is no other option than using multiple TCP connections between each pair of source-destination.
% (e.g, one TCP connection for the data of each single object or one TCP connection for the data of a subset of objects).
Either way, the design of TCP does not allow the~transport layer and~for~the~network to identify the flows that belong to the same application and that are transferring data of all objects/sensors that contribute to run the same application (e.g.,~imagine the example a virtual environment built out from different objects located in~different geographic location). 
%What if multiple points (nodes) need to communicate
%----------------------------------------------------------------
%-------------------------------------------------------------------------------------------------
%$\bullet$ \textbf{Unawareness of the application's flows:} As mentioned earlier, in today's Internet transport layer is that an application might use multiple TCP connections to transport data related to different objects. However, 
As~a~result, the transport layer and the network operate and manage these flows independently from each other. 

For futuristic applications, a flow belonging to an application should be managed while taking into account (1)~the~existence of~other transport flows belonging to the same application (even if they are not originating from the~~same~~source), and~(2)~the~application-level performance requirements for each of these flows, which may vary over time. For instance, depending on the application context, the throughput of some connections might be decreased to allow to increase other flows' throughput and vice-versa. Packet dropping decisions might be also taken while considering the changing performance requirements of these flows and their associated priorities.

$\bullet$ \textbf{Blind congestion control:}
%----------------------------------------------------------------
In current networks, TCP tries to blindly guess the state of the network in order to~attempt to~avoid congestion. In particular, TCP uses two strategies,~timeouts and~duplicate acknowledgment, to detect potential packet loss. However, these strategies are not accurate and may wrongly assume that~some packets are lost. 
This may lead to unneeded packet retransmissions and inappropriate decisions from the TCP congestion control scheme that may reduce data transmission rate without being really needed. 
The lack of accurate information at~the~transport layer regarding the state of~network is~definitely a~major limitation of current TCP-based transport layer. 

%----------------------------------------------
$\bullet$ \textbf{Retransmission delays:} 
%----------------------------------------------
As the transport layer is responsible for the end-to-end communication between end points, retransmission of lost packets is carried out by the source end point. 
As a result, a significant additional delay is experienced whenever a packet is lost.
For instance, using TCP, this delay would be equal to the timeout or the time needed to receive three duplicate packets added to the end-to-end delay from the source to~the~destination. This additional delay is~roughly equal to one round trip time plus the end-to-end delay\footnote{The round trip time is the time needed to send a~packet to~the~destination and receive its acknowledgment. The end-to-end delay is the time to send the~packet from the~source to~the~destination}. 
It is clear that the way reliability is implemented by~TCP is~a~handicap for~many flows that require ultra-low latency. 

%-------------------------------------------------
$\bullet$ \textbf{Few words about QUIC and SCTP:} 
%-------------------------------------------------
%
%-------------------------------------------------------------------------------
%\subsubsection{QUIC and SCTP Limitations} \label{sec:QUICandSCTPLimitations}
%-------------------------------------------------------------------------------
%The additive-increase/multiplicative-decrease (AIMD)
%When needed, SCTP fragments user messages to ensure that the SCTP packet passed to the lower layer conforms to the path MTU.  On receipt, fragments are reassembled into complete messages before being passed to the SCTP user  \ref{SCTP}.
%
Although TCP is the most used transport protocol today, new transport protocols like SCTP and QUIC have been recently proposed to address some of the TCP's limitations. In this section, we describe how these protocols work and highlight the issues that they are still not able to solve.
 
The Stream Control Transmission Protocol (SCTP) is a connection-oriented transport protocol that offers a~reliable full-duplex association that refers to~a~communication between exactly two systems (an~``association'' is~equivalent to~a~``connection'' in~TCP)~\cite{SCTP-rfc4960}. Unlike~TCP that assumes a single stream of data, an~SCTP association transfers simultaneously multiple independent streams of~messages between the two end points over the same session. Each stream is~a~sequence of~messages that should be delivered %in the correct sequence (i.e.,~order) 
to~the upper-layer protocol at~the~receiving side. A~key~advantage of~SCTP~over~TCP~is~that~a~packet loss may block the delivery of only the stream to which it belongs, whereas other streams are delivered normally to the upper layer without any delay. SCTP could also allow partial reliability by indicating the~level~of reliability for each message \cite{SCTP-rfc3758-Reliability}.
%and the maximal number of retransmissions that should be considered for that message \cite{SCTP-rfc3758-Reliability}.

% Reliability https://tools.ietf.org/html/rfc3758#section-4.1
%[RFC3758] 1. how the service user indicates the level of reliability required for a particular message, and
%[RFC3758] 2. how the sender side implementation uses that reliability level to determine when to give up on further retransmissions of that message.
			
Another transport protocol that have been recently proposed is QUIC (Quick UDP Internet Connections) \cite{QUIC-UDP,QUIC-HTTP,quic-protocol-02}. 
QUIC's~main goal~is~to~improve the~performance of http traffic in terms of latency and bandwidth. 
It~allows to~establish a~connection between two endpoints on top of UDP where a single connection multiplexes several streams of data.
In~addition~to stream multiplexing, QUIC provides a credit-based flow control similar to TCP where the receiver advertises the number of~bytes it could accept for each stream and for the entire connection~\cite{quic-protocol-02}. QUIC~also allows the~application to indicate the~relative priority of the streams in~order to~decide which data will be transmitted first.

It is also worth noting that both, SCTP and QUIC, implement congestion control schemes that operates like the ones of~TCP (i.e.,~slow-start and~congestion avoidance algorithms) \cite{quic-recovery-16,SCTP-rfc4960}. 
Unlike TCP, both protocols allows multi-homing (i.e.,~each endpoint could support multiple IP addresses at~the~same~time so there is no need to instantiate another connection if the IP address changes). However, multi-homing is only used for redundancy purposes (i.e.,~not~for~load balancing). 

\begin{table}[!tb]
\centering
\caption{Comparison of the services offered by TCP, QUIC and SCTP}
\label{TB:TCPvsOthers}
\begin{tabular}{|l|l|l|l|}
\hline
\multicolumn{1}{|c|}{\textbf{Service}}    & \multicolumn{1}{c|}{\textbf{TCP}}                                                                & \multicolumn{1}{c|}{\textbf{QUIC}}                                                                  & \multicolumn{1}{c|}{\textbf{SCTP}}                                                                                    \\ \hline
\begin{tabular}[c]{@{}l@{}}\textbf{Communication} \\ \textbf{type} \end{tabular} & \begin{tabular}[c]{@{}l@{}}Two end points \\ (one single flow)\end{tabular} & \begin{tabular}[c]{@{}l@{}}Two end points \\ (multiple streams)\end{tabular} & \begin{tabular}[c]{@{}l@{}} Two end points\\  called association\\(multiple streams) \end{tabular} \\ \hline
\textbf{Flow control}                     & Yes & Similar to TCP & Similar to TCP\\ \hline
\begin{tabular}[c]{@{}l@{}}\textbf{Congestion}\\ \textbf{Control}\end{tabular}              & \begin{tabular}[c]{@{}l@{}}Slow-start \\+ Cong. avoidance\end{tabular} & Similar to TCP & Similar to TCP \\ \hline
\textbf{Muli-streaming}                   & Not supported  & Yes & Yes      \\ \hline
%\textbf{Number of streams}                   & Not supported  & Variable during a connection & Fixed at the association's initialization      \\ \hline
%\textbf{Stream lifetime}                   & Not supported  & Different for each stream & Same as the association's lifetime      \\ \hline
\begin{tabular}[c]{@{}l@{}}\textbf{Stream initiation}\\ \textbf{/teardown}\end{tabular}                    & Not supported  & \begin{tabular}[c]{@{}l@{}}At any time during\\ the communication\end{tabular} & \begin{tabular}[c]{@{}l@{}}Only at initiation\\ /teardown \end{tabular}  \\ \hline
\begin{tabular}[c]{@{}l@{}}\textbf{Stream} \\ \textbf{flow control} \end{tabular} & Not supported  & Yes  & \begin{tabular}[c]{@{}l@{}}Combined\\ (Not per stream)\end{tabular}  \\ \hline
%https://tools.ietf.org/html/draft-joseph-quic-comparison-quic-sctp-00#section-4.6
\begin{tabular}[c]{@{}l@{}}\textbf{Stream} \\ \textbf{prioritization}    \end{tabular}        & Not supported & Yes & Yes   \\ \hline
\textbf{Stream reliability}               &  \begin{tabular}[c]{@{}l@{}} Single flow\end{tabular}
  & Supported  & \begin{tabular}[c]{@{}l@{}}Supported \end{tabular}   \\ \hline
\textbf{Ordered delivery}         & \begin{tabular}[c]{@{}l@{}}Within a flow\\(mandatory)\end{tabular}                              & Within a stream& \begin{tabular}[c]{@{}l@{}}Within a stream \\ (optional)\end{tabular} \\ \hline
\begin{tabular}[c]{@{}l@{}}\textbf{Explicit congestion}\\ \textbf{notification}\end{tabular}  & Optional & Optional  & Optional \\ \hline
\textbf{Multi-homing}  & Not Supported  & Yes   & Yes  \\ \hline
\textbf{Reliability} & \begin{tabular}[c]{@{}l@{}}Mandatory \\for all packets \end{tabular}      & Full/Partial  & Full/Partial  \\ \hline
\end{tabular}
\end{table}

Table~\ref{TB:TCPvsOthers} compares services and features offered by TCP, QUIC, SCTP \cite{QUIC-UDP,QUIC-HTTP,quic-protocol-02,SCTP-rfc4960,joseph-quic-comparison-quic-sctp-00}. 
Compared to TCP, QUIC and SCTP provide more multi-streaming, multi-homing and more flexibility in terms of reliability. However, both still support only connections between two end~points. As~a~result, the transport layer is not aware of all the flows (or streams) that belong to the same application when more than two end-points are involved. This does not allow the transport layer to~take into account~all~the~streams when carrying~out flow and~congestion control. 
Furthermore, both QUIC and TCP still use the~same congestion control scheme of TCP and hence inherent its limitations. That is, they have to go through the slow start (i.e.,~gradual increase of the sending rate) and congestion avoidance (i.e.,~linear increase of the sending rate). They~are~not~also able to~accurately detect packet loss as they rely on timeouts and duplicate acknowledgment.
%(the same one used by~TCP) that is not

Finally, the support of the network to these transport protocols is very limited. Indeed, the network does not notify the~transport layer or support it in performing its services like flow and congestion control and packet retransmission. The~only potential support is~the~Explicit Congestion Notification (ECN) that provides some hints to the transport protocol to~react to~congestion. 
However, ECN~does not~provide accurate information about the location and gravity of the congestion. 
In~addition, ECN~notifications are sent to all TCP connections going through the same congested router, which may lead them~to~reduce all their sending traffic at~the~same time. It~is~clear that a~better collaboration and more exchange of information between the transport and network layers may potentially address these issues. 
\subsection{Network Layer Protocols} \label{sec:NetworkLayer}
%-------------------------------------------------------------------------------

Current network layer protocols mainly offer ``best effort'' data delivery by routing traffic towards its destinations.  They~do not~provide any~performance guarantees in terms of throughput, delay and packet loss.
Furthermore, the network layer is not aware of the applications' composition in terms of flows (i.e.,~the~flows belonging to~the~same application), their~performance requirements, and how these requirements are changing over time. 
As a result, it could not dynamically adjust the routing strategy and adapt the congestion control according to these requirements.

Furthermore, even when a congestion happens, the network drops packets ``blindly'' without taking into consideration their~priority, their belonging flows and their relevance for the application. 
Another important limitation is that current networks have only three options to react to congestion: dropping packets, balancing the load across multiple paths or~dynamically changing the route. However, these three options may impact the application performance. Packet dropping incurs additional delays for retransmission. Balancing the load may lead to overload some paths and underuse others. Changing routing paths assumes that~there are paths that have enough bandwidth to~carry the~traffic, which may not be always true.

Finally, in current TCP/IP protocol stack, the network and the transport layers do not collaborate to address the congestion in the network. Each layer operates independently of the other and tries to reduce congestion with the limited knowledge it~has. 
For instance, the network layer does not provide any explicit feedback about the packet loss and the congestion state in~the~network and does not support the transport layer to~control flow rates or to ensure data reliability and retransmitting lost data.

%----------------------------------------------------------
\subsection{Network Stack Headers} \label{sec:StackHeaders1}
%----------------------------------------------------------
Another major limitation of today's Internet is that the current protocols use headers that do not provide, by default, meta-data and information that will be mandatory to manage the traffic of future applications. These headers are~also~not~customizable enough to be adjusted to the application and its associated traffic engineering requirements. Ideally, depending on the needs of~the application, the packet header should include meta-data with information about objects/sensors, flows (e.g.,~flow ids and their ownership to~application), applications' requirements, type and relevance of the carried data.

Furthermore, the current protocols are also not flexible enough to incorporate commands that could be executed by~the~network like requests for a particular route or~packet processing (e.g.,~data compression). 
Finally, current protocols have several fields in the headers that are rarely used or might be useless in many cases. For instance, MAC addresses in the Ethernet header are~not~relevant for point to point communication and the Type of Service field in the IP header is not used in most cases.
%Recent proposals advocated to include additional meta -data and commands to the IP header (e.g.,~Keynote of Dr Richard Li, Huawei USA). This additional information will allow application to convey information about the application’s flows as well as routing and performance requirements to the resource management framework and to the forwarding nodes (including VNFs) as well. 
%In this context, a challenging task is to adopt this new IP layer and identify the metadata that should be included to the IP header and how it shall be used by the network in order to satisfy the requirements of  futuristic applications.

%----------------------------------------------------------
%\subsection{SDN/NFV paradigm} 
%----------------------------------------------------------

%----------------------------------------------------------
\subsection{Sources of Latency} 
%----------------------------------------------------------
%----------------------------------------------
%$\bullet$ \textbf{Sources of delay}:
%----------------------------------------------
As one of the requirements of future applications is ultra-low latency, we identify and analyze in this Section the different sources of delays in the network.
Table~\ref{TB:SrcDelay} defines these different delays and provide the parameters on which they depend. 
Of course, in order to reduce the end-to-end delay, its composing delays should be~shrunk.
We can note that media-dependent delays, like the transmission delay, require further improvements to~existing media and to the functions of the physical layer (e.g.,~coding) in~order to increase their transmission capacity. 
All the other delays (e.g.,~retransmission and steering delay) are governed by the network architecture and protocols. As a result, future network architecture and protocols should provide efficient solutions to accurately control these delays.

It is also worth noting that, while edge computing is a promising solution to reduce delays in the network, it may not be beneficial for all applications, especially when users involved in the same communication are geographically far from each other. In this case, the traffic has to cross the Internet to connect these users, and hence all the aforementioned delays have to~be~experienced.

%\begin{table}[!htb]
%\caption{Source of delays in the network}
%\label{TB:SrcDelay}

\begin{table}[!htb]
\centering
\caption{Sources of delay in the network}
\label{TB:SrcDelay}
\begin{tabular}{|l|l|lll}
\cline{1-2}
\textbf{Source of Delay}      & \multicolumn{1}{c|}{\textbf{Comment}}                                                                                                                                                                                                                                                  &  &  &  \\ \cline{1-2}
\textbf{Propagation delay}    & \begin{tabular}[c]{@{}l@{}}It refers to the time needed for a signal to go through \\ transmission media. This time depends on the characteristics \\of the media (i.e., signal propagation speed and distance)\end{tabular}                                                                   &  &  &  \\ \cline{1-2}
\textbf{Transmission delay}   & \begin{tabular}[c]{@{}l@{}}It is the time needed to inject a packet into the media. \\ It depends on the transmission capacity of the media (bps)\\ and the packet size.\end{tabular}                                                                                                    &  &  &  \\ \cline{1-2}
\textbf{Processing delay}     & \begin{tabular}[c]{@{}l@{}}It is defined as the time needed by a network function \\or an application to process a packet.\\ It is usually at the scale of nanoseconds for hardware-based\\ network functions but can be much higher \\for software-based network functions\end{tabular}  &  &  &  \\ \cline{1-2}
\textbf{Queuing delay}        & \begin{tabular}[c]{@{}l@{}}It is the time that a packet waits for in a network component \\(device or network function) before it can be processed \\ or transmitted. It depends mainly on the load of the network. \\ A high queuing delay is a sign of congestion in the network device.\end{tabular}                      &  &  &  \\ \cline{1-2}
\textbf{Retransmission delay} & \begin{tabular}[c]{@{}l@{}}It is the time needed to detect a packet loss and retransmit it. \\A retransmission may be triggered by the link layer (e.g., Wi-Fi) \\ and also by the transport layer when TCP is used. In this case, \\this delay is at least 3 times the end-to-end delay.\end{tabular} &  &  &  \\ \cline{1-2}
\textbf{Steering delay} & \begin{tabular}[c]{@{}l@{}}It is the time incurred when the traffic is first steered to a location\\ to be processed before it is forwarded to the final destination.\end{tabular} &  &  &  \\ \cline{1-2}
\textbf{End-to-end delay} & \begin{tabular}[c]{@{}l@{}}It is the total time needed to send a packet from the source \\to the destination. It encompasses all the delays above.\end{tabular} &  &  &  \\ \cline{1-2}
\end{tabular}
\end{table}

%=================================================================================
\section{FlexNGIA: Flexible Next-Generation Internet Architecture} \label{sec:FlexNGIA}
%=================================================================================
In what follows, we describe the proposed Flexible Next-Generation Internet Architecture adapted to the Tactile Internet (FlexNGIA), and provide its key architectural elements in terms of infrastructure, services, management framework, network protocol stack and headers, and~describe the~potential business model adapted to this new architecture.
Throughout this Section, we discuss also the advantages of~this~new architecture and how it overcomes the limitations of today's Internet.
% We start first by describing future network infrastructures and components. We then propose first a new network architecture that is more adapted to typical future applications.  We also put forward a management framework to allocate and manage the resources of the physical infrastructure.

%---------------------------------------------------------------------
\subsection{Network Infrastructure}\label{sec:NetworkInfrastructure}
%--------------------------------------------------------------------

%----------------------------------------------------
\begin{figure}[htbp]
	\centering
		\includegraphics[width=0.99\textwidth]{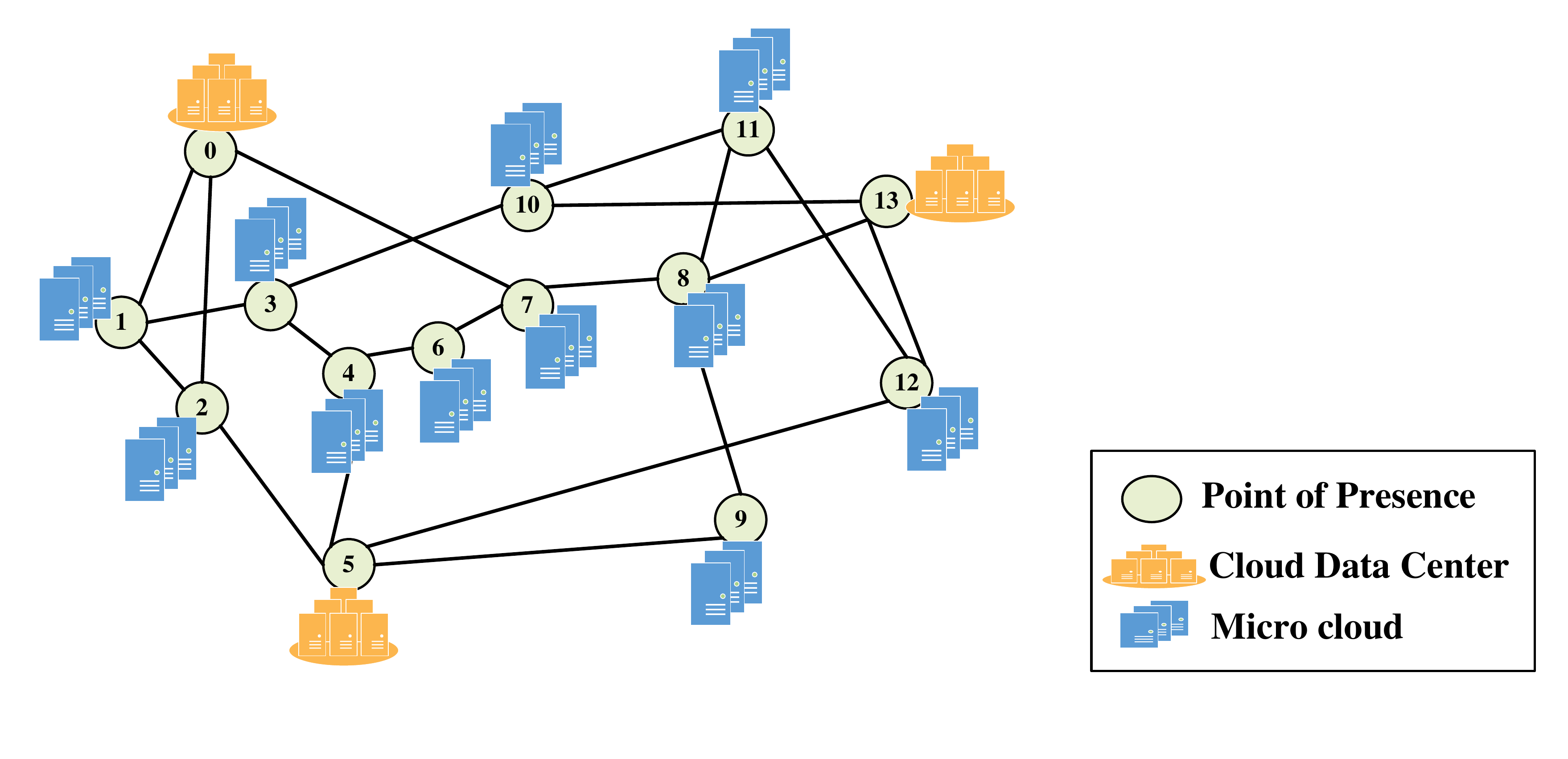}
	\caption{Network infrastructure}
	\label{fig:PhysicalInfrastructure}
\end{figure}
%----------------------------------------------------

The main characteristic of future network infrastructures is the availability of fully programmable computing and networking resources not only at the edge of the network but also at the core as shown in Fig.~\ref{fig:PhysicalInfrastructure}. The core nodes of the network are Points of Presence (POPs) that will consist of routers, switches and also computing resources (e.g.,~servers, GPU, NPU). 
%
%This allows them to have full control over their infrastructures and to customize them. 
The main features of future infrastructures could be summarized as~follows:  

$\bullet$ \textbf{Edge to edge infrastructures:} today's cloud providers are building infrastructures everywhere from the edge of~the~network to~the~core. Such infrastructures are made out from cloud data centers with large amounts of resources and~several micro clouds with potentially less computing resources towards the edge \cite{BariCST13,Elkhatib2017Fog,OLANIYAN2018}.

%It is clear that there may not be possible to have a~large amount of resources everywhere like in cloud data centers but the edge should have the minimum required resources to support applications (i.e., micro clouds in Fig.~\ref{fig:PhysicalInfrastructure}).
%It is clear that there may not be possible to have a~large amount of resources everywhere like in cloud data centers but the edge should have the minimum required resources to support applications (i.e., micro clouds in Fig.~\ref{fig:PhysicalInfrastructure}).

$\bullet$	\textbf{Computing resources are everywhere:} in future infrastructures, computing resources should be available throughout the~infrastructure including the network core and the network edge. However, it is worth noting that the amount of computing resources may be variable depending on the location. These resources could be powerful servers with advanced processing capabilities like Field-Programmable Gate Arrays (FPGAs) and Network Processing Units (NPUs) or simply commodity servers with basic capabilities. Of course, such equipment allows to run any network function or task in any node of the network. This allows \textit{In-Network Computing}, which is a major change compared to traditional networks. This provides the advantage of eliminating any steering delay by processing data in the network nodes that are in the path towards the destination.

$\bullet$ \textbf{Programmability and advanced network functions:} this feature is realized through several technologies including, but~not~limited to, software defined networking and network function virtualization. Software defined networking allows to~separate the network control plane from the data plane. It advocates implementing the control plane as a logically centralized controller that provides switches with the forwarding rules to handle the traffic. At the same time, network function virtualization advocates to run traditional network functions (e.g.,~routing, firewall, intrusion detection) on virtual machines or containers hosted in commodity servers. While these technologies radically change the way the network is configured and its functions are provisioned, they can be used to go beyond these simple changes. Indeed, future network functions should go beyond the~traditional ones (i.e., packet forwarding) to include more advanced functions like video compression and encoding, congestion control, application data aggregation and processing.

%------------------------------------------------------------------------------
\begin{figure} [!t]
	\centering
		\includegraphics[width=0.80\textwidth]{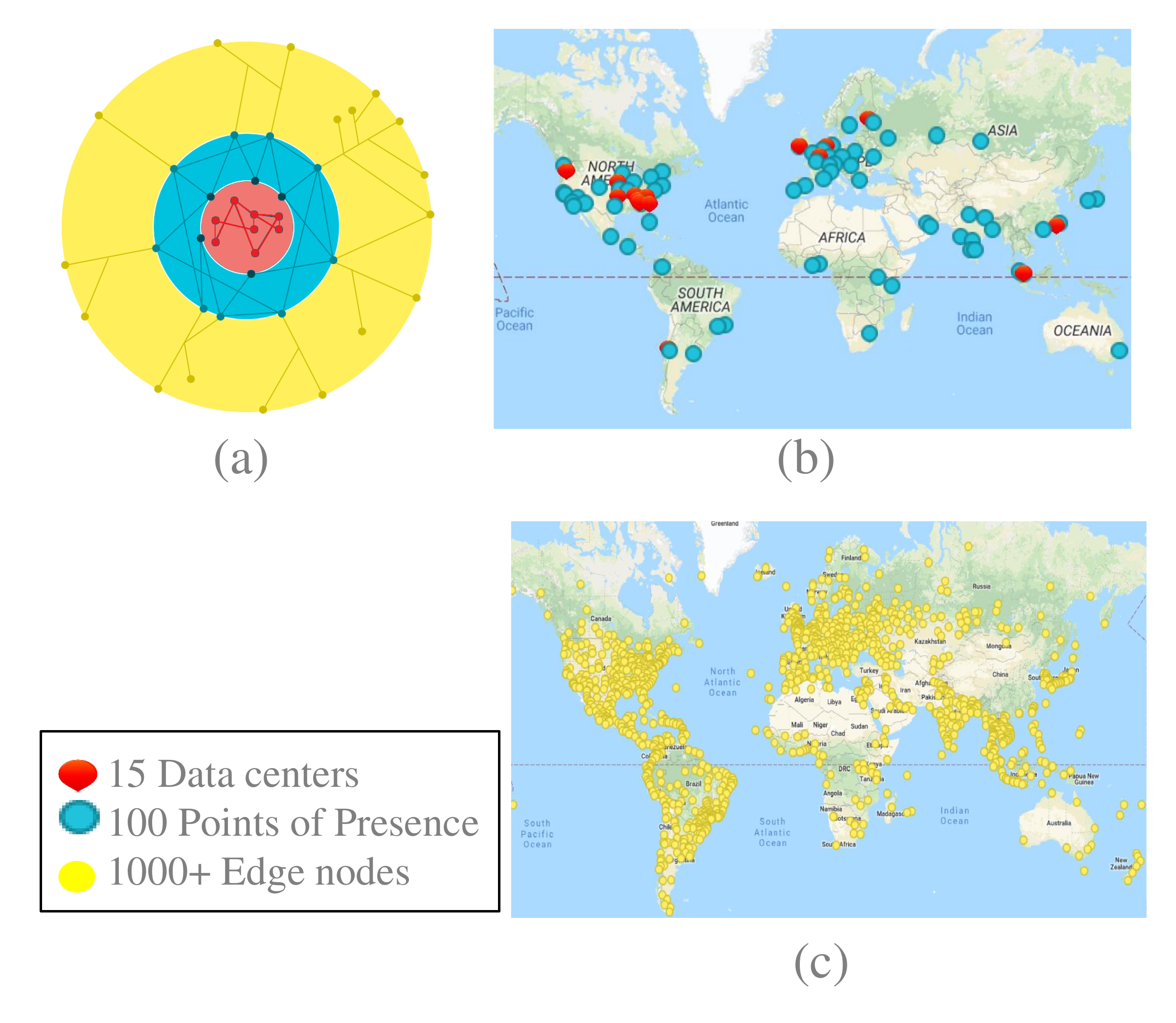}
	\caption{Example of an edge to edge infrastructure: Google Cloud Platform (Source:\cite{GoogleInfrastructure})}
	\label{fig:GoogleInternet}
\end{figure}
%--------------------------------------------------------------------------------

It is clear that today's trend is that cloud providers and companies build and use their own private transit backbones rather than public transit backbone. Several web-scale companies such as Google, Facebook and Amazon have already started to~build their edge to edge private infrastructures. 
For~instance, Fig.~\ref{fig:GoogleInternet} shows the Google Cloud Platform, which is a world-wide software defined platform. The infrastructure is made out from 15 data centers, 100 points of presences and more than 1000 edge nodes equipped with computing resources \cite{GoogleInfrastructure}.

A key advantage of having such a private world-wide network is that the infrastructure's owner has full  control over~all the~resources from the edge to the edge, and hence, has the full flexibility and ability to program these resources and configure the network and the computing resources in order to be better guarantee the sought-after performance objectives.

%==================================================================
\subsection{Network Services} \label{sec:NetworkServicesFunctions}
%==================================================================

%--------------------------------------------------------------------------------
\begin{figure}[htb]
	\centering
		\includegraphics[width=0.80\textwidth]{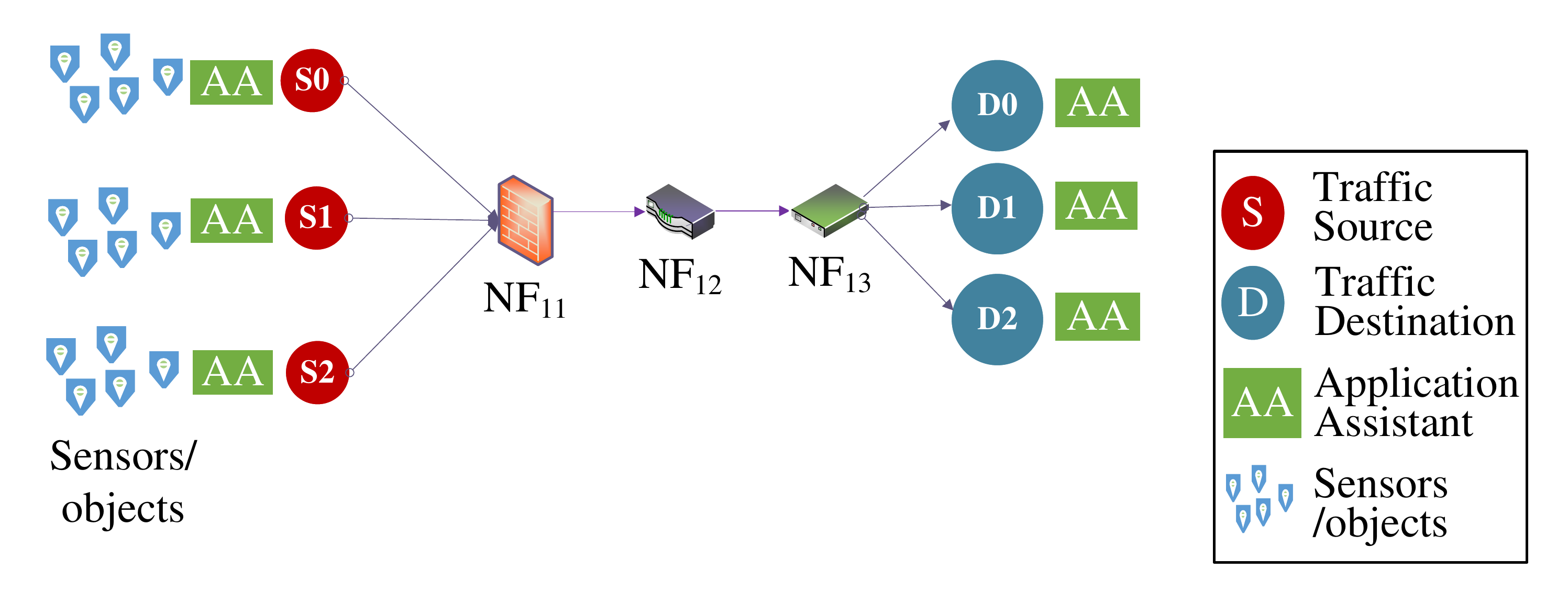}
	\caption{Example of a Service Function Chain Associated with an Application}
	\label{fig:SFC}
\end{figure}
%--------------------------------------------------------------------------------
%Service Function Chain SFC1 associated with Application 1

Future networks should offer not only data delivery service but also advanced services like reliability and application-tailored services. Thanks to technologies like virtualization and software defined networking, the network will allow to dynamically provision \textit{Service Function Chains} (SFCs) tailored to each application. As~shown in Fig.~\ref{fig:SFC}, an SFC is made out from \textit{Virtual Network Functions} (VNFs) connected through virtual links. Service function chains and their composing VNFs could be tailored based on the application's need and performance requirements. As~shown in the figure, a service function chain supports data transmission from multiple sources towards multiple destinations, which allows to capture the octopus-like characteristic of future applications connecting multiple users at~the~same~time (e.g.,~the~use case of~``virtual coffee shop'' described in~Section~\ref{sec:RequirementsAGlanceIntoTheFuture}).

As seen in the figure, at each endpoint (i.e.,~source or destination), a~special network function called \textit{Application Assistant (AA)} is deployed to~receives the incoming or the outgoing traffic from the sources and destinations, respectively. The~AA operates at the application layer and is in charge of to aggregating the~traffic coming from different sensors/objects, to process it, tag it or filter it based on the application requirements and~context (e.g.,~requested performance, data relevance, flow priority).

When the service function chain is designed, one should define the following elements: 
\begin{itemize}
	\item Sources and destinations of the traffic
	\item The types of each of its composing network functions, its implementation, its input packet format and its output packet format, and its requirements in terms of hardware and resources (e.g., GPU, NPU, CPU, memory and disk) and performance (e.g., packet processing rate, processing delay).
	\item The overall requested performance of the chain (e.g., end-to-end delay, packet processing rate)
	\item The communication protocols that should be deployed to ensure the communication between the end-points of the service function chain and its composing network functions. FlexNGIA allows to design and deploy any communication protocol above layer 2. More details on how this is achieved are provided in subsection \ref{sec:StackHeaders}.
\end{itemize}

We provided above some elements defining a service function chain associated with an application. However, this definition could be expanded to include more elements depending on the context and the application requirements.

\textbf{How this definition is a game-changing for the Internet landscape and industry?} FlexNGIA provides the designer of the chain/application the flexibility of designing not only the end points (i.e., client and server of an application) but also the communication protocols and the network functions that are deployed in the network. We therefore do not restrict the communication protocols to the common ones (e.g., TCP, UDP, QUIC, IPv4 or IPv6) but opening the door to the application designers to further innovate and develop customized network functions and communication protocols at the different layers that are adapted to their applications (layer 3 and above). We can therefore imagine in a near future an application store (similar to Google Play Store or the Apple App Store) where different applications are available with different network functions and protocol stack. When the application is launched, the whole associated service function chain including its network functions and communication protocols are dynamically provisioned in the network.
This is definitely a game-changing concept that opens the door for limitless innovation for future Internet applications with customized communication protocols and network services.

In the following, we provide more details about the network functions that could compose the service chain as advocated by FlexNGIA. 
%are provided in~the~next subsection.
%~\ref{sec:NetworkProtocolStackArchitecture}.

%
%For instance, the network may support the transport layer to ensure the reliability of data transfer and to solve congestion problems in order to satisfy performance guarantees. 

%==============================================================================================
\subsection{Network Functions and Protocol Stack} \label{sec:NetworkProtocolStackArchitecture}
%==============================================================================================
%\subsubsection{FlexNGIA Protocol Stack and Network functions} \label{sec:FlexNGIAProtocolStackNetworkFunc}
%$\bullet$ \textbf{FlexNGIA Protocol Stack and Network functions}

%------------------------------------------------------------------
\begin{figure}[!bp]
	\centering
		\includegraphics[width=0.90\textwidth]{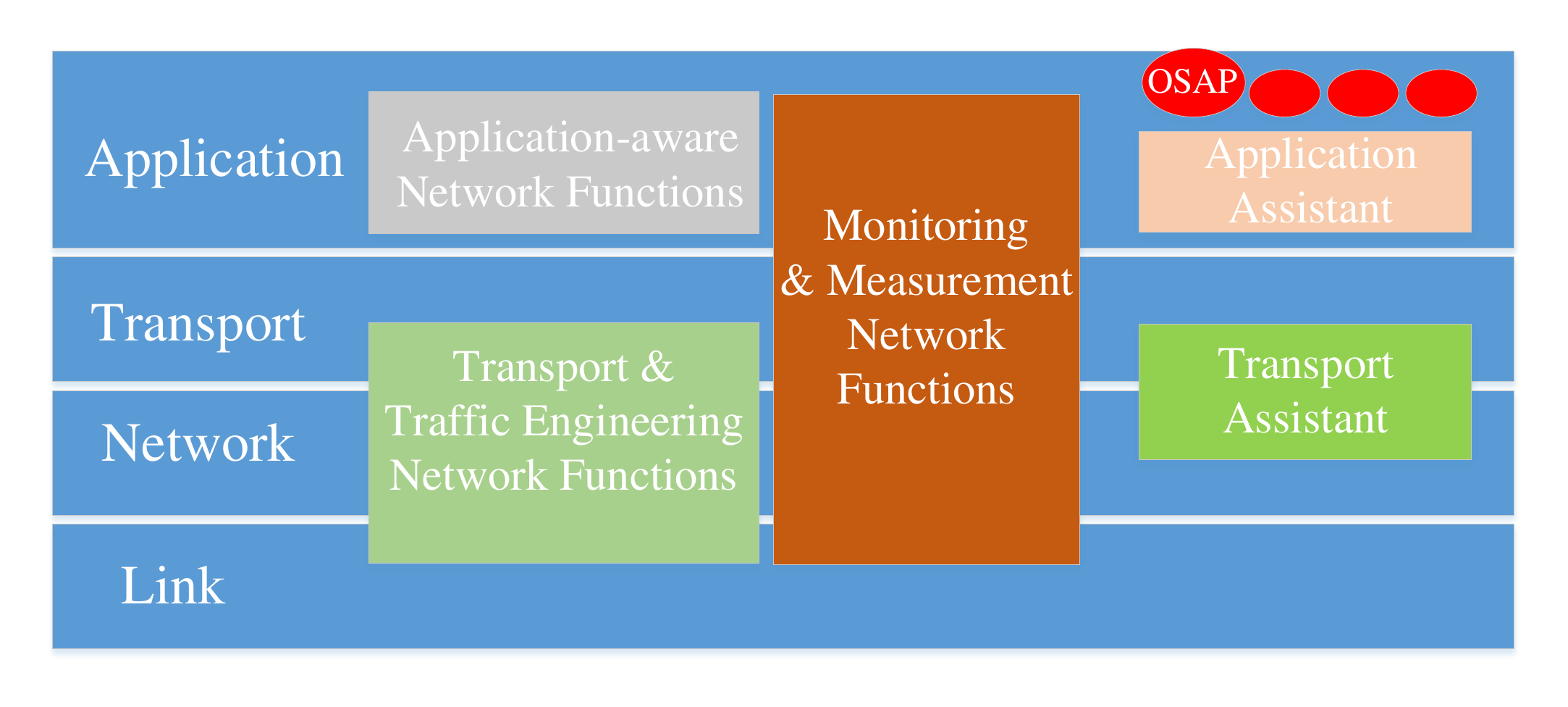}
	\caption{Classes of advanced functions of FlexNGIA and their operating layers. The figure shows also two special network functions proposed in FlexNGIA (i.e., Application assistant and Transport Assistant)}
	\label{fig:OSAP}
\end{figure}
%------------------------------------------------------------------

%The remaining network functions in the chain could be of any type. In what follows, 
With FlexNGIA, we~define~two~types of functions that~could be provided in~future networks:

%The functions are not limited to traditional network functions (e.g.,~firewall, intrusion detection) but also include advanced functions like packet retransmission, data processing, compression, object rendering, subtitling, incorporating data and objects into flows (e.g.,~augmented reality).

$\bullet$ \textbf{Basic network functions:} The network will definitely offer data delivery through traditional (or basic) network functions like routing, packet forwarding, firewalls and intrusion detection systems.

$\bullet$ \textbf{Advanced network functions:} 
As future networks contain computing resources throughout the infrastructure (e.g.,~commodity servers, dedicated network hardware, FPGA, GPU, NPU), they will be able to carry out in-network computing to support applications. As a result, depending on the application, advanced application-aware functions could be deployed dynamically in~the~network to perform data processing, machine learning, data compression, object rendering, stream multiplexing, incorporating data into flows (e.g.,~subtitling for video, merging or adding objects for augmented reality) and caching. Such functions could operate at different layers (e.g.,~OSI~layer~1 to~layer 7 or~even at~the~queue management level). 
As~a~result, the~concept of~software-defined networking should go beyond configuring forwarding elements (i.e.,~forwarding rules) to allow to  dynamically configure these advanced functions (e.g.,~how selective reject should work, what kind of processing the network should apply to a video streaming).

%\caption{Examples of application-aware traffic management functions}
%\label{ExamplesAdvancedFunctions}

\begin{table}[!b]
\centering
\caption{Examples of advanced network functions}
\label{ExamplesAdvancedFunctions}
\begin{tabular}{lll}
\hline
\multicolumn{1}{|c|}{\textbf{Function}}      & \multicolumn{1}{c|}{\textbf{Description}} & \multicolumn{1}{c|}{\textbf{Benefit}}\\ \hline
%-----------------------------------------------------------------------------------
\multicolumn{3}{|l|}{\textbf{Application-Aware Functions}} \\ \hline
%-----------------------------------------------------------------------------------
%\multicolumn{1}{|l|}{Content grouping} & \multicolumn{1}{l|}{\begin{tabular}[c]{@{}l@{}}Several packets are grouped into a single one to reduce the header size \\or to combine their content.\end{tabular}} & \multicolumn{1}{l|}{Reduce traffic size}\\ \hline
\multicolumn{1}{|l|}{\begin{tabular}[c]{@{}l@{}}Data \\processing\end{tabular}}       & \multicolumn{1}{l|}{\begin{tabular}[c]{@{}l@{}} Processing the content of incoming packets \\(e.g.,~application-level processing like \\ data compression, video cropping)\end{tabular}} & \multicolumn{1}{l|}{\begin{tabular}[c]{@{}l@{}}Support application \\ Reduce traffic size \\ (data compression)\end{tabular}} \\ \hline
\multicolumn{1}{|l|}{\begin{tabular}[c]{@{}l@{}}Flow\\multiplexing\end{tabular}} & \multicolumn{1}{l|}{\begin{tabular}[c]{@{}l@{}}Merge multiple flows based on the content \\ (e.g., subtitling for video, incorporating data, \\ merging or adding objects for augmented reality)\end{tabular}} & \multicolumn{1}{l|}{\begin{tabular}[c]{@{}l@{}}Support application\end{tabular}} \\ \hline  
%-----------------------------------------------------------------------------------
%\multicolumn{3}{|l|}{\textbf{Switching Functions}} \\ \hline
%-----------------------------------------------------------------------------------
%\multicolumn{1}{|l|}{Function duplication}   & \multicolumn{1}{l|}{Duplicate a function in the network if needed.} & \multicolumn{1}{l|}{Reduce congestion} \\ \hline
%-----------------------------------------------------------------------------------
\multicolumn{3}{|l|}{\textbf{Transport and Traffic Engineering Functions }} \\ \hline
\multicolumn{1}{|l|}{Forwarding}    & \multicolumn{1}{l|}{Forward packets according to forwarding rules} & \multicolumn{1}{l|}{\begin{tabular}[c]{@{}l@{}}Reduce congestion \\ and delay\end{tabular}}  \\ \hline
\multicolumn{1}{|l|}{Load balacing}  & \multicolumn{1}{l|}{Distribute packets across several paths} & \multicolumn{1}{l|}{\begin{tabular}[c]{@{}l@{}}Reduce congestion \\and delay\end{tabular}} \\ \hline
%-----------------------------------------------------------------------------------
%\multicolumn{1}{|l|}{Packet caching}         & \multicolumn{1}{l|}{\begin{tabular}[c]{@{}l@{}}Cache packets in this network functions in order to retransmit them if a loss \\ is detected. This allows faster retransmission delays\end{tabular}}  & \multicolumn{1}{l|}{\begin{tabular}[c]{@{}l@{}}Ensure reliability\\ Reduce delay\end{tabular}} \\ \hline
\multicolumn{1}{|l|}{\begin{tabular}[c]{@{}l@{}}Packet caching \\\& retransmission\end{tabular}} & \multicolumn{1}{l|}{\begin{tabular}[c]{@{}l@{}}Cache packets in the network to retransmit them\\ if a loss is detected. This allows faster retransmis-\\sion delays (see Transport Assistant in Sect.~\ref{sec:NetworkProtocolStackArchitecture})\end{tabular}}  & \multicolumn{1}{l|}{\begin{tabular}[c]{@{}l@{}}Ensure reliability\\ Reduce delay\end{tabular}} \\ \hline
\multicolumn{1}{|l|}{\begin{tabular}[c]{@{}l@{}}Congestion \\control\end{tabular}} & \multicolumn{1}{l|}{\begin{tabular}[c]{@{}l@{}}Store packets in order to gradually transmit them \\ at a different sending rate. This function could use \\ slow start and congestion avoidance algorithms \\ or novel congestion control algorithms adapted \\ to the application\end{tabular}}  & \multicolumn{1}{l|}{\begin{tabular}[c]{@{}l@{}}Reduce congestion\end{tabular}} \\ \hline
\multicolumn{1}{|l|}{\begin{tabular}[c]{@{}l@{}}Packet\\ duplication\end{tabular}} & \multicolumn{1}{l|}{\begin{tabular}[c]{@{}l@{}}Duplicate relevant packets through different paths \\ to ensure packets are received at the destination\end{tabular}} & \multicolumn{1}{l|}{Ensure reliability} \\ \hline  
\multicolumn{1}{|l|}{\begin{tabular}[c]{@{}l@{}}Packet\\ bundling\end{tabular}}        & \multicolumn{1}{l|}{\begin{tabular}[c]{@{}l@{}}Several packets are grouped into a single one to\\ reduce the header size or to combine their content\end{tabular}} & \multicolumn{1}{l|}{Reduce traffic size}\\ \hline
\multicolumn{1}{|l|}{\begin{tabular}[c]{@{}l@{}}Selective \\ packet drop\end{tabular}}  & \multicolumn{1}{l|}{\begin{tabular}[c]{@{}l@{}}Packets are dropped selectively based on priority\\ or application's context and requirements\end{tabular}}        & \multicolumn{1}{l|}{Reduce congestion} \\ \hline
\multicolumn{1}{|l|}{\begin{tabular}[c]{@{}l@{}}Partial \\ packet drop\end{tabular}}  & \multicolumn{1}{l|}{\begin{tabular}[c]{@{}l@{}}
Some bytes are dropped from the packet if this\\ will not impact the user's quality of experience\end{tabular}}        & \multicolumn{1}{l|}{Reduce congestion} \\ \hline   
%
%
%\multicolumn{1}{|l|}{Selective packet drop}  & \multicolumn{2}{c|}{\textbf{One Column}} \\ \hline
%-----------------------------------------------------------------------------------
\multicolumn{3}{|l|}{\textbf{Monitoring  \& Measurements Functions}} \\ \hline
%-----------------------------------------------------------------------------------
\multicolumn{1}{|l|}{\begin{tabular}[c]{@{}l@{}}Coarse-grained \\ monitoring\end{tabular}} & \multicolumn{1}{l|}{\begin{tabular}[c]{@{}l@{}} Collect coarse-grained statistics about the flows \\ (e.g., average queuing delay, average processing \\delay, rate)\end{tabular}} & \multicolumn{1}{l|}{\begin{tabular}[c]{@{}l@{}}Monitoring\end{tabular}} \\ \hline  
\multicolumn{1}{|l|}{\begin{tabular}[c]{@{}l@{}}Per-packet \\ monitoring\end{tabular}} & \multicolumn{1}{l|}{\begin{tabular}[c]{@{}l@{}} Incorporate monitoring information within \\ the packet header (e.g., timestamp, \\queuing delay, processing delay)\end{tabular}} & \multicolumn{1}{l|}{\begin{tabular}[c]{@{}l@{}}High-precision\\ monitoring\end{tabular}} \\ \hline 
%\multicolumn{1}{|l|}{Flow multiplexing} & \multicolumn{1}{l|}{\begin{tabular}[c]{@{}l@{}} Merge multiple flows based on the content (e.g., subtitling for video, incorporating data, \\merging or adding objects for augmented reality)\end{tabular}} & \multicolumn{1}{l|}{Reduce traffic\\leverage in-network computing to avoid steering delays}\\ \hline  
                                                                                             
\end{tabular}
\end{table}

Fig.~\ref{fig:OSAP} shows the three classes of network functions defined in FlexNGIA and their operating layer in the protocol stack architecture. Table~\ref{ExamplesAdvancedFunctions} shows these classes with some examples of functions and describes their~goal and potential benefits to improve application and~network performance and to support the operation of~application, transport and network layers. Specifically, the three classes of~functions are as follows:

\begin{itemize}

	\item \textit{Application-aware network functions:} These functions operate at the application layer and~are tailored to each application. They are hence aware of the application' logic and characteristics like~traffic flows used by~the~same application (even if they are connecting multiple sources and multiple destinations) and the requirements of~each of these flows in terms of performance, reliability and~availability.
They should be also aware of~the~type of~the~transmitted data and the application context. They~can hence process packets or take decisions (e.g.,~selectively drop packets, compress data) taking into account the application requirements and~user experience. The Application Assistant introduced above could be considered as an example of~an~application-level network function.

%\textit{- Switching network functions:} 
	\item \textit{Transport and traffic engineering network functions:} These functions could provide advanced forwarding and load balancing functions (e.g.,~rule-based forwarding and routing). They also offer mainly transport layer services (e.g.,~congestion control, packet caching and retransmission) and traffic engineering functions (e.g.,~selective packet dropping). FlexNGIA defines a~special network function called \textit{Transport Assistant (TA)} that implements such functions. More details about the TA are~provided in~the following paragraphs. %subsection~\ref{sec:NetworkProtocolStackArchitecture}.

	\item \textit{Monitoring and measurements network functions:} These functions aim at monitoring and measuring the performance metrics of the network services and functions.

\end{itemize}

FlexNGIA proposes also two special network functions, called Application Assistant and Transport Assistant, that could significantly support future network applications. The use of these functions is not mandatory and any other network function could be imagined, designed and proposed. The way they are implemented and their operation could be customized depending on the application's requirement. Fig.~\ref{fig:OSAP} shows their operating layer and shows also a new proposed concept called Object Service Access Point (OSAP) allowing to handle multiple sensors at the network edge. In the following, we provide more details about the OSAP and these two functions:

%We do not constraint their implementation details as they can be customized .we provide in the following the layers at which FlexNGIA network functions operate with respect to today's Internet protocol stack (Fig.~\ref{fig:OSAP}):

%$-$ 
\begin{itemize}
	\item \textsl{Object Service Access Point (OSAP):} an object service access point is an~end point for~communication between an~object (or a sensor) and~the~application assistant. Different objects or sensors are connected through the OSAPs to the application assistant which is the first network function in the  service function chain associated with the application. 
	\item \textsl{Application Assistant (AA):} an application assistant is instantiated at each of the chain endpoints. An endpoint at~a~particular chain corresponds to a location of a user (sources or destination). The application assistant is in charge of receiving and controlling data~flows arriving from several objects or~sensors at this endpoint and then forwards them to the following functions in~the~service function chain. 

The application assistant is also responsible for setting the application requirements in terms of performance and~reliability for each of the flows coming from the objects/sensors. It can also dynamically configure the objects or~sensors (e.g.,~data throughput, compression rate) depending on the application requirements, context and~user behavior. For~instance, in~an~augmented-reality application, a~user might have several virtual objects superimposed in~his~view. However, in~practice, he may focus on using only one of these objects. In~this~case, the~application assistant identifies such behavior and~reduces the~throughput of the data associated to~the~less used objects. Furthermore, the application assistant conveys the~application's requirements (i.e.,~throughput, flow priority, requested reliability) to the network functions of the chain and potentially (when~needed) to~the~application controller located at the central resource management. The application assistant incorporates such information in the packet headers as metadata or~commands. It is hence in charge of mapping the application context and requirements into metadata incorporated into the packets.
	
%In~the~aforementioned applications, several objects will be connected to~an~application assistant. For~instance, in~a~telepresence application, the traffic of multiple 3D objects is directed to the application assistant to be rendered and~then~transmitted to one or multiple users (i.e.,~destinations).
%(Fig.~\ref{fig:ManagementFramework})

%, i.e.,~the~transport assistant associated with this particular application assistant. 
%It is worth noting that~there~is~an~application assistant at each end-point of~a~particular~application.
%The application assistant conveys application's requirements (i.e.,~throughput, flow priority, requested reliability) along with the received data flows to the following network functions and to the application controller located at the central resource management.

%$\bullet$	\textbf{Application assistant:} as mentioned previously, application assistants are instantiated for each application at each end-point. An application assistant operates at the application layer and 
%%It receives several data flows from different objects and~sensors and~dynamically estimates the  requirements of each flow (i.e.,~throughput, flow priority, requested reliability) and feeds it to the transport assistant that maps these in the network.
%it is in charge of receiving and controlling data~flows arriving from several objects or sensors. 

%$-$ 
\item \textsl{Transport Assistant (TA):} the transport assistant is a cross-layer component that carries out functions of the traditional transport layer (e.g.,~TCP) and network layer (e.g.,~IP) at the same time (Fig.~\ref{fig:OSAP}). 
%As shown in Fig.~\ref{fig:OSAP}, the transport assistant is a cross-layer component that spans the transport and network layers. 
It can be seen as an advanced network function that is part of the provisioned service function chain. Hence, it can be instantiated many times in the chain throughout the path to~the~destinations depending on the needs of~the~application.
The transport assistant receives the data flows and also information about the performance requirements of each of them from the~application assistant. These requirements could be different from one object flow to another and could also change over time. The transport assistant is responsible for congestion control, data loss management, packet retransmission (see~Table~\ref{ExamplesAdvancedFunctions} for~more~examples). These services are offered while taking into account all the flows of the same application and their requirements in terms of performance and reliability. 
\end{itemize}

In the following, we provide more details about the benefits of the TA and the combination of the transport and network layers.\\

%is aware of the flows belonging to the same application. 

%$\bullet$ \textbf{Congestion control:} 
%-------------------------------------
%==============================================================================================
\subsection{FlexNGIA Cross-layer Transport} \label{sec:FlexNGIACross-layerTransport}
%==============================================================================================
%$\bullet$ \textbf{FlexNGIA Cross-layer Transport}\\
%---------------------------------------------------------------------------------------------------
\begin{figure}[!b]
	\centering
		\includegraphics[width=0.99\textwidth]{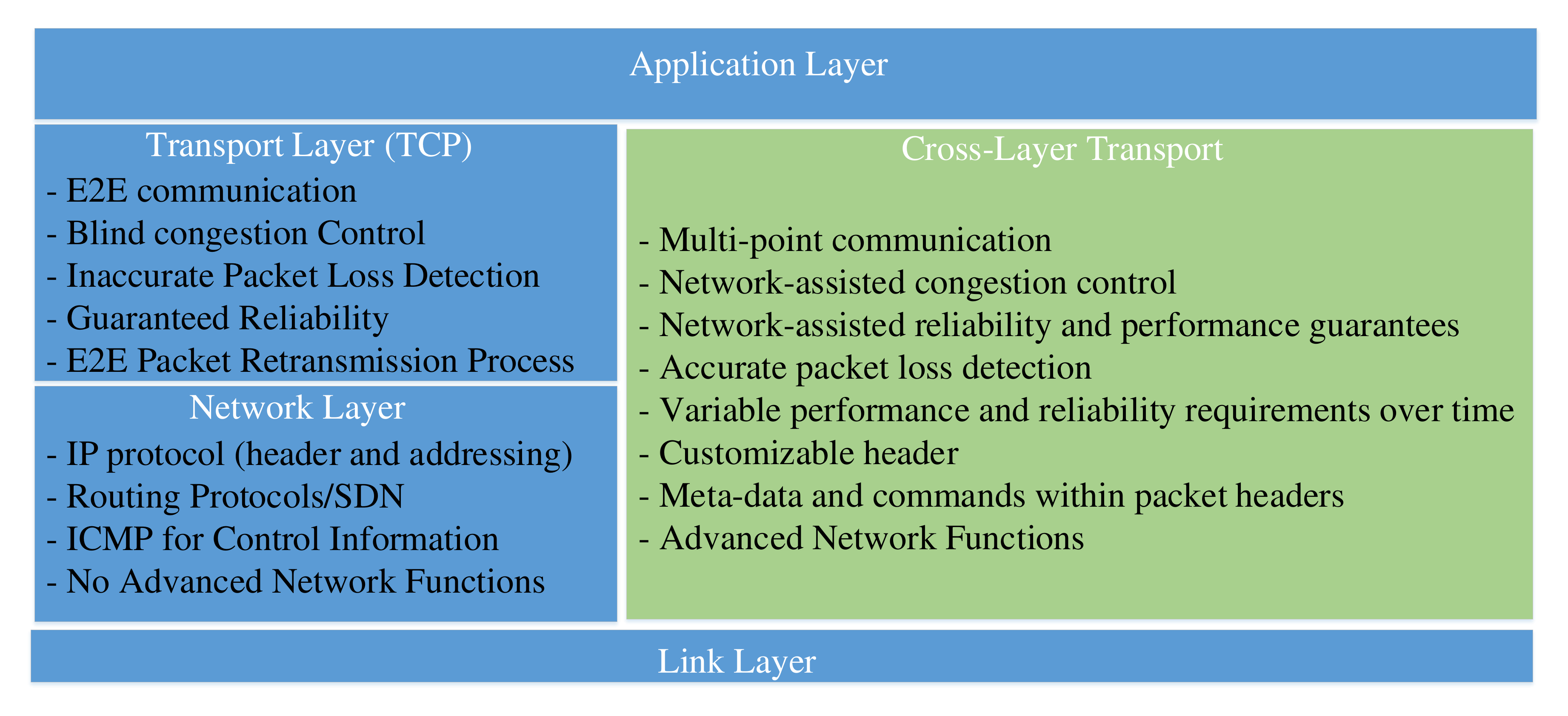}
	\caption{Proposed network stack architecture (with cross-layer transport) compared to the TCP/IP architecture}
	\label{fig:CrossLayerArchitecture}
\end{figure}
%---------------------------------------------------------------------------------------------------

One of the main features advocated by FlexNGIA (though not mandatory) is to merge the traditional transport and network layers to~build a~cross-layer  transport layer that combines both layers. 
Fig.~\ref{fig:CrossLayerArchitecture} compares the~functions of the traditional layers to~that~of~the~proposed cross-layer transport. 
%the main benefit of such layer combination is that flow congestion control is done by a single entity and that the network is aware of the flows belonging to the same application. 
As shown in the figure, the~main benefit of~such layer combination is that layer 3 and 4 services are performed by a single entity (i.e.,~the~combined~layer).

Unlike traditional networks where congestion control and traffic routing problems are addressed separately at two different layers (i.e.,~transport layer and the network layer), the proposed cross-layer transport solves these problems altogether taking into account the applications' requirements (provided by the application assistant) and the network state.
For instance, when there is congestion in the network, the combined layer will make sure that, depending on the application's requirements, context and user behavior (captured by the application assistant), some flows are delivered with stringent performance and reliability requirements while others could tolerate low quality of service. This could be done by dynamically adjusting the amount of~dropped packets from each flow, caching packets, readjusting the data sending rate, adjusting the packet retransmission rate, changing routing paths, and by adjusting the number, size and location of~the~virtual network functions that are processing packets from these flows.

Another benefit of this cross-layer design is that the network could also offer other advanced functions like data reliability that can be achieved through in-network packet caching and retransmission originated from the network (rather from the source endpoint) as the network is aware of the application's and flow requirements at run-time (see~Table~\ref{ExamplesAdvancedFunctions} for more examples of~advanced functions). 

%\textcolor{blue}{As these in-network advanced network functions that implement the combined layer proposed in FlexNGIA could implement application-specific features and transport layer services. It is worth noting that the proposed combined layer in FlexNGIA breaks the end-to-end paradigm of the traditional transport layer defined in the OSIAs these advanced network functions running within the network could implement application-specific features and transport layer services that reside in the endpoints model and used in current application and~transport protocols (e.g.,~UDP, TCP, QUIC, SCTP).}

It is worth noting that the proposed combined layer in FlexNGIA breaks the end-to-end principle adopted by today's Internet and existing application and~transport protocols (e.g.,~UDP, TCP, QUIC, SCTP). Indeed, in FlexNGIA, application-specific features and transport layer services are not  implemented only in the end points but could be also offered or supported by~advanced functions (e.g.,~transport assistants) running within the network.

%Indeed, in FlexNGIA, application-specific features and transport layer services could be offered or supported by advanced functions running within the network and not only in the endpoints.
%In FlexNGIA, the proposed in-network advanced functions could implement application-specific features and transport layer services which usually reside in the endpoints.
%As discussed earlier, advanced network functions running within the network could implement application-specific features and transport layer services that reside in the endpoints.

%--------------------------------------------------------------------------------------
\begin{figure}[htbp]
	\centering
		\includegraphics[width=0.80\textwidth]{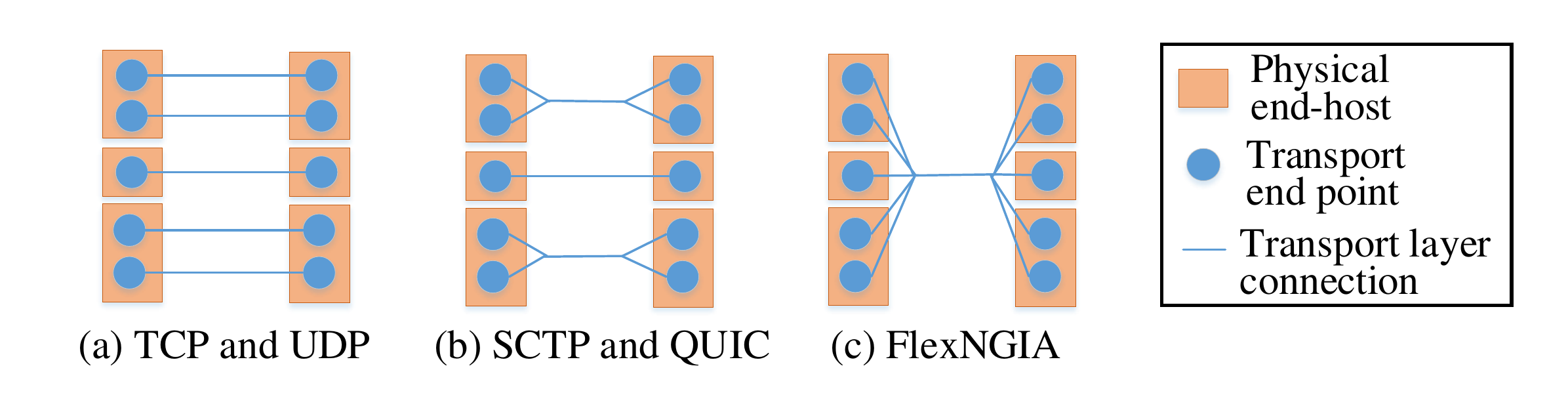}
	\caption{Illustration of how one single network application might be seen at the transport layer}
	\label{fig:TransportConnections}
\end{figure}
%--------------------------------------------------------------------------------------

To better highlight the benefits of the cross-layer transport design, let us assume that we have one single application that~involves several physical hosts and several objects (e.g.,~virtual reality objects). This could be the case of a virtual reality environment built out from different objects transferred from different locations.
Fig.~\ref{fig:TransportConnections} shows how this application is seen by~the~transport layer in~traditional networks using TCP, SCTP and QUIC compared to how it is seen with FlexNGIA's cross-layer transport.
As~shown~in~Fig.~\ref{fig:TransportConnections}~(a), the transport layer using TCP is not aware of the flows belonging to the same application (even when these flows have the same source and destination) and hence each end-to-end TCP communication is handled separately from the others. 
With SCTP and QUIC (Fig.~\ref{fig:TransportConnections}~(b)),  the transport layer becomes aware of the flows (or~streams) of the same application when they connect the same source and destination hosts. 
The transport layer manages then each flow (or stream) while taking into account the others. However, when these flows connect different source and destination hosts, the transport layer and the network handle them in a separate manner without considering that they all belong to~the~same application. 
With the cross-layer design of~FlexNGIA (Fig.~\ref{fig:TransportConnections}~(c)), all the end-points involved in the same application are identified and the traffic shared among them is recognized by the network and is associated to the same application (thanks to the information added to the packet headers and to the awareness of the management framework and the chain network functions). This allows the network and transport layers to be fully aware of the application composition in terms of flows and~requirements.
 
As~a~result, the~transport assistants could manage all these flows while taking into account that they all belong to the same application. They monitor these flows and divide the total bandwidth allocated for the application among them. They could also control the congestion and route the traffic while considering the type and meaning of the data carried by each of the flows and its importance for the application (depending on the context and~user behavior). For instance, if a virtual reality user is manipulating a particular virtual object, the flow carrying the data of this object becomes more important than that~of~the~other flows. The transport assistants at the end points or within the network will react accordingly.

%\textcolor{blue}{Furthermore, the service function chain deployed for the application is allocated specific amounts of bandwidth for each of the virtual links connecting the network functions. The transport assistants within the chain are in charge of dynamically controlling the proportion of bandwidth used by each of the flows belonging to the same application. }
%if the application were allocated If an application requires X bandwidth, then at any point in time, the total bandwidth that can be allocated to all application flows should not exceed X. Does the transport assistant keep track of the bandwidth used by all flows that belong to the same application ?

Please note that Fig.~\ref{fig:TransportConnections} shows an abstraction of the application's communications at the transport layer. This means that it~does not~show the real paths taken by~the traffic between the different end-points (i.e.,~packets between the~end-points do not necessarily take the~same path).

%----------------------------------------------------------
\subsection{Network Stack Headers} \label{sec:StackHeaders}
%----------------------------------------------------------
%We need the following: 
%In general, packet headers should be flexible enough to include two types of information in addition to the application data. These information are meta-data which 
As FlexNGIA relies on a logically centralized management framework and service function chains to carry applications' traffic, we~are~defining the two following types of packets to carry management and application data:

$-$ \textsl{Signaling packets:} these packets do not carry data and are used mainly to instantiate an~application and to convey its initial requirements (i.e., SFC specifications) in terms of network functions (types of network functions, their resource requirements in terms of cpu, memory, disk) as well as the virtual links connecting these functions (e.g.,~bandwidth, delay).

$-$  \textsl{Regular packets:} these packets carry mainly the application data (if there are any) in addition to metadata (i.e.,~additional data pertaining to the application or any other function of the network layer stack) and also commands that could be executed by the network functions and the application control module. Table~\ref{ExamplesMetadataandCommands} provides examples of possible metadata and commands. Depending on the application, different types of metadata and commands could be defined.
\begin{table}[!b]
\centering
\caption{Examples of Metadata and commands}
\label{ExamplesMetadataandCommands}
\begin{tabular}{|l|l|}
\hline
\multicolumn{1}{|c|}{\textbf{Example of Metadata}}
& \multicolumn{1}{c|}{\textbf{Examples of  commands}}
\\ \hline
\begin{tabular}[c]{@{}l@{}}- Application id \\ - Object/sensor source \\  - Object/sensor destination(s)\\ - Packet Priority\\ - Performance requirements at run-time \\(e.g.,~throughput, delay, packet loss)\\ - Retransmission needed or not\\ - Retransmission timeout (i.e., retransmission \\ is not needed after this time out)\\ - Routing preference\\ - Video/audio compression rate, requirements, \\and quality\\ - Timestamp at particular nodes \\ - Packet processing time at a particular nodes \\- Packet queuing delay at particular nodes 
\end{tabular} & \begin{tabular}[c]{@{}l@{}}- Drop if congestion\\ - Compress video to a minimum \\ of X bps if needed\\ - Crop the video carried by this flow \\ to a particular surface if needed\\ - Send only to user X and Y\\ - Add timestamp \\ - Add values of packet processing\\ and queuing delay \\ if they exceed a threshold\\ - Add delay value from source \\ - Send report if this packet's delay\\ exceeds a threshold \\ \\ 
\end{tabular} 
\\ \hline
\end{tabular}
\end{table}
Furthermore, the format of regular packets should be as follows:\\
1) A layer-2 header that contains mainly an \textit{application id} (similar to a VLAN id) which is used to identify a single application and to carry out packet forwarding through the application SFC. This is similar to having a layer-2 network virtualization where a virtual network is considered to connect the network functions needed for a single the application. 

This virtualization layer provides several advantages. First, the application id could encode the operator id that initiated the application (on-behalf of the user) in addition to a unique identifier for the application itself within the operator network. Consequently, the~resulting application id could be kept the same throughout different networks of different operators. This addresses the drawbacks of VLAN technology \cite{VLAN}, which provides a limited number of possible VLANs ids that could not be kept throughout). This also removes the need for other technologies like VXLAN \cite{rfc7348VXLAN}, which adds additional encapsultations (MAC in UDP) or like Overlay Transport Virtualization (OTV) \cite{OTV} as the application ID could be kept when the packet goes through different networks and data centers. \\\\
%
% 2 mac addresses-> 12 bytes*8 bits= 96 bits
% 32 bits ->2^32 operators -> 4 billion operators.
% 64 bits -> 2^32 application id -> 
% today's TCP+UDP number of connections -> 32+32+16+16+2= 96 bits but the Average number of TCP/UDP connections per application is high. 
% type ID of Ethernet = FlexNGIA
%
2) The upper-layer headers of the packet are flexible headers that could include metadata and commands defined by~the~application designer. Of~course, the network functions in the chain that will process the application traffic should be aware of~the~expected format.  
The designer of the application decides of the packet format (i.e., header~and payload) of the packets expected and~generated by the application assistants and by each network function in the chain.

This fully flexible header format provides the possibility to tailor the packets to the needs of each~application and the network functions of the service chain that will be deployed to steer and process the application's traffic.

%The upper-layer headers will include the metadata and the commands that will be defined depending on the application requirements. 

%---------------------------------------------------------------------
\subsection{Network Management Framework}\label{sec:NetworkManagement}
%---------------------------------------------------------------------

%--------------------------------------------------------------------------------
\begin{figure}[th!]
	\centering
		\includegraphics[width=0.90\textwidth]{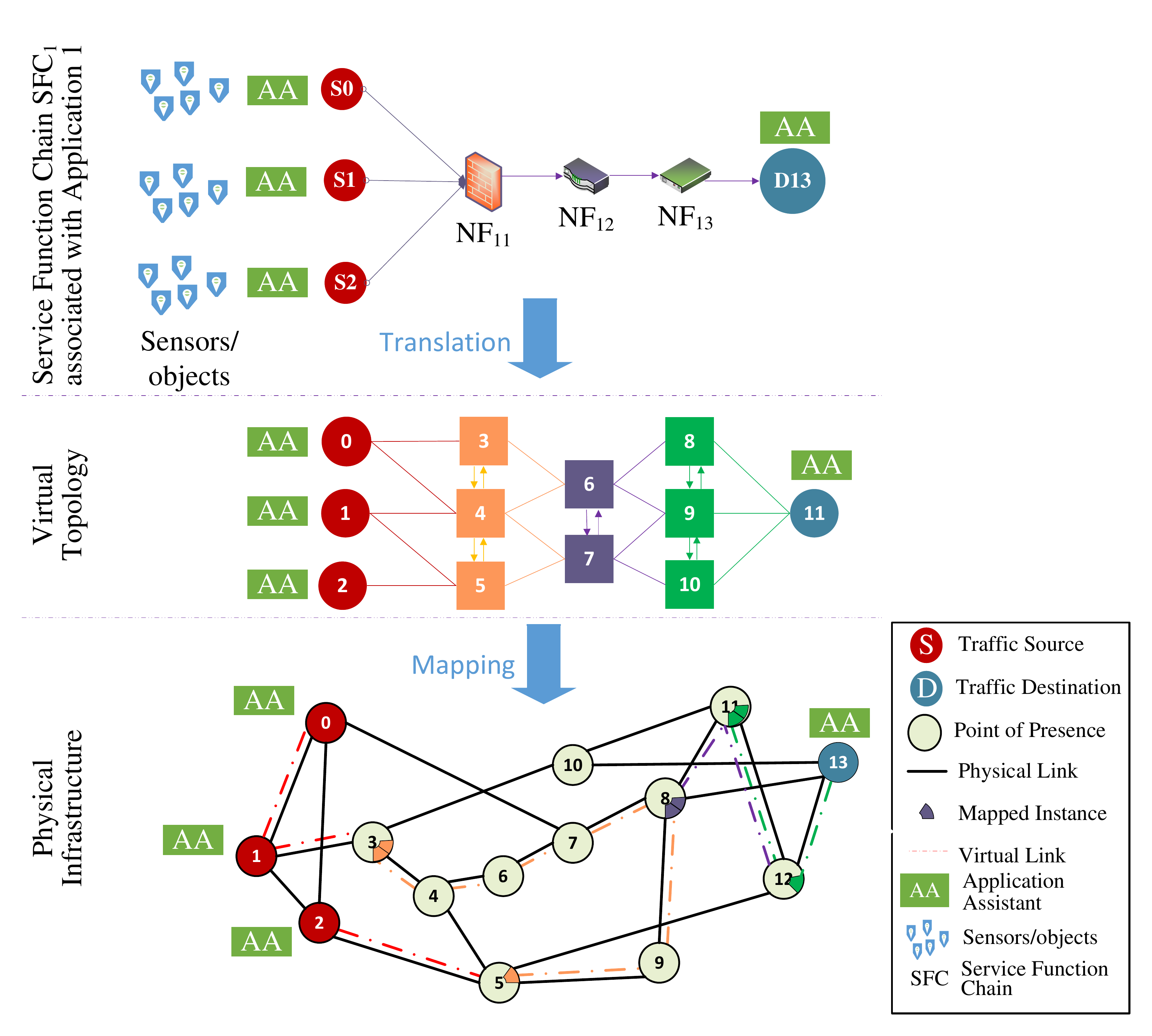}
	\caption{Example of Resource Allocation for a Service Function Chain}
	\label{fig:SFCEmbedding}
\end{figure}
%--------------------------------------------------------------------------------

Before delving into the details of the FlexNGIA resource management framework, we first describe an example of~how a~service function chain associated with an application could be allocated resources.

\textbf{$\bullet$ Resource Allocation for Service Function Chains}

Fig.~\ref{fig:SFCEmbedding} shows how a service function chain is allocated computing and networking resources. The resource allocation of the chain is carried out in mainly two steps: 1) Translation and 2) Mapping. A service function chain associated with an application is first translated into a~virtual topology. To build this virtual topology, each network function is translated into a certain number of instances. %contains instances of the network function composing the service chain. 
An instance could be either a~container or~a~virtual machine implementing a~specific network function. 
The~number and~size of~the~instances for each network function depend on several parameters like~the~demand (i.e.,~incoming traffic), the~type of~the~function, the~instance processing capacity (e.g.,~number of processed packets per second) and~cost. The~work of~Ghrada~et~al.,~\cite{Ghrada-PVESDN2018} provides some insights on how to select virtual machine instances based~on~the~network function type, processing capacity and~cost.

The instances are connected through virtual links such that the traffic goes through the functions in the order defined in the chain. Each virtual link has specific requirements in terms of bandwidth and~propagation delay. Even the instances implementing the same function may need to~synchronize and share some data pertaining to the operation of the function itself, and~hence may need to~be~connected with~virtual links. For~example,~instances implementing an~intrusion detection system may need to synchronize information about potential attacks. 

Once the virtual topology is defined, the next step is the mapping onto the physical infrastructure. In this step, resources are~allocated for all instances and virtual links while taking into consideration their performance requirements. Other~objectives may be also considered during the mapping step like~the~cost minimization, energy efficiency, availability of green sources of~energy. 

In the example described in~Fig.~\ref{fig:SFCEmbedding}, network functions $NF_{11}$, $NF_{12}$ and~$NF_{13}$ require three, two and three instances, respectively.  The resulting virtual topology is then mapped into the infrastructure. For example, instances 3 and~4 are~mapped onto POP 3 while instance 5 is~mapped onto POP~5, respectively.

\textbf{$\bullet$ FlexNGIA Management Framework}

%--------------------------------------------------------------------------------
\begin{figure} [!t]
	\centering
		\includegraphics[width=0.99\textwidth]{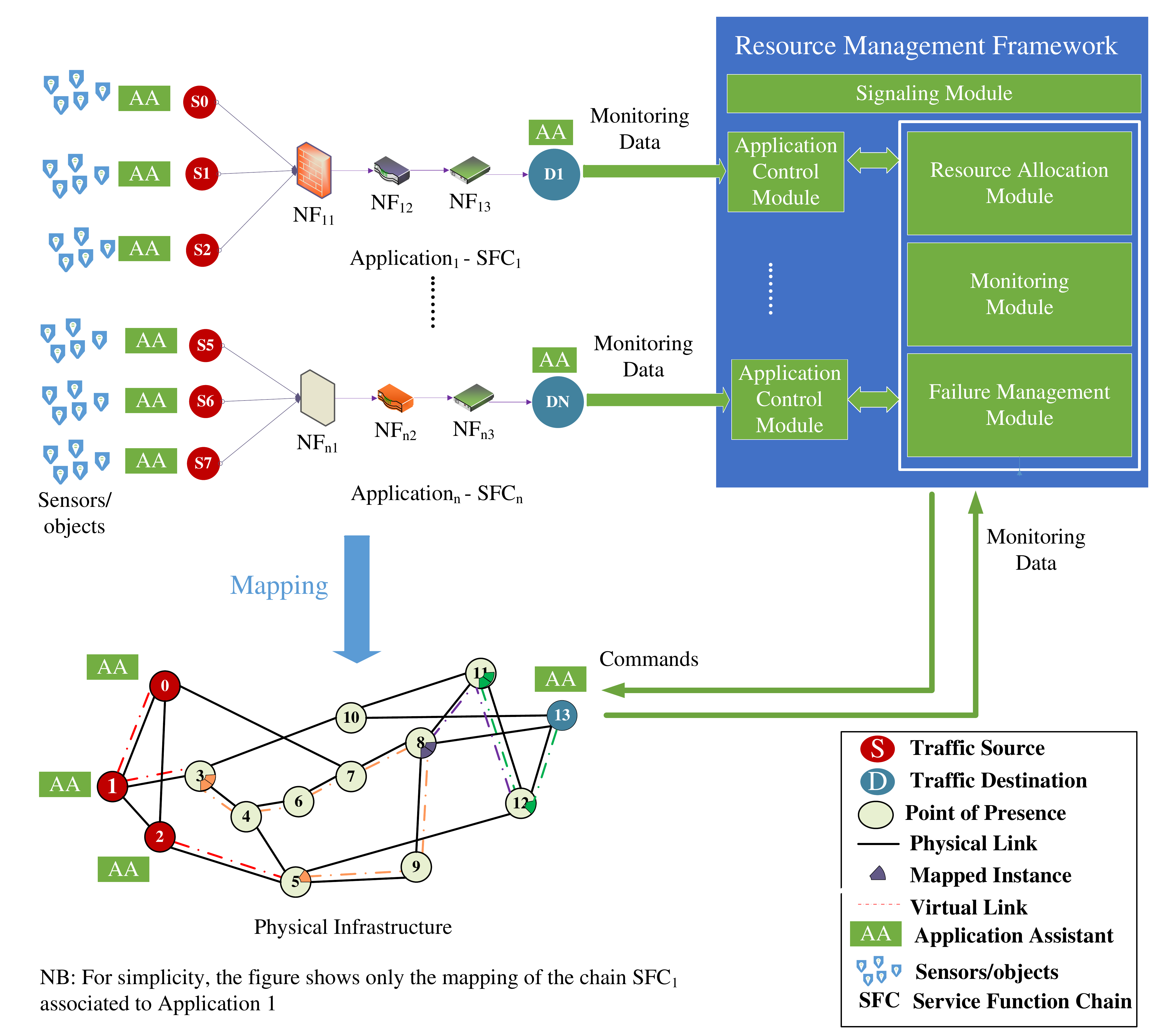}
	\caption{Network Management Framework}
	\label{fig:ManagementFramework}
\end{figure}
%--------------------------------------------------------------------------------

As future network operators do not only offer packet delivery but service function chains, there is~a~compelling need to~define a more sophisticated resource management framework to allocate and manage the resources of these chains while, at~the~same~time, ensuring their performance, reliability and availability requirements and achieving the network operators' objectives (e.g.,~energy efficiency, usage of green sources of energy, load balancing). 

As shown in Fig.~\ref{fig:ManagementFramework}, %shows the FlexNGIA central resource management framework that is supposed to manage the infrastructure and~to~map service function chains onto the physical resources. The 
the FlexNGIA management framework is a~logically-centralized framework that manages the resources of~the~whole infrastructure. 
In the following, we describe the role of the modules composing~this~framework:

%--------------------------------------------------------------------------------
\begin{figure}
	\centering
		\includegraphics[width=0.75\textwidth]{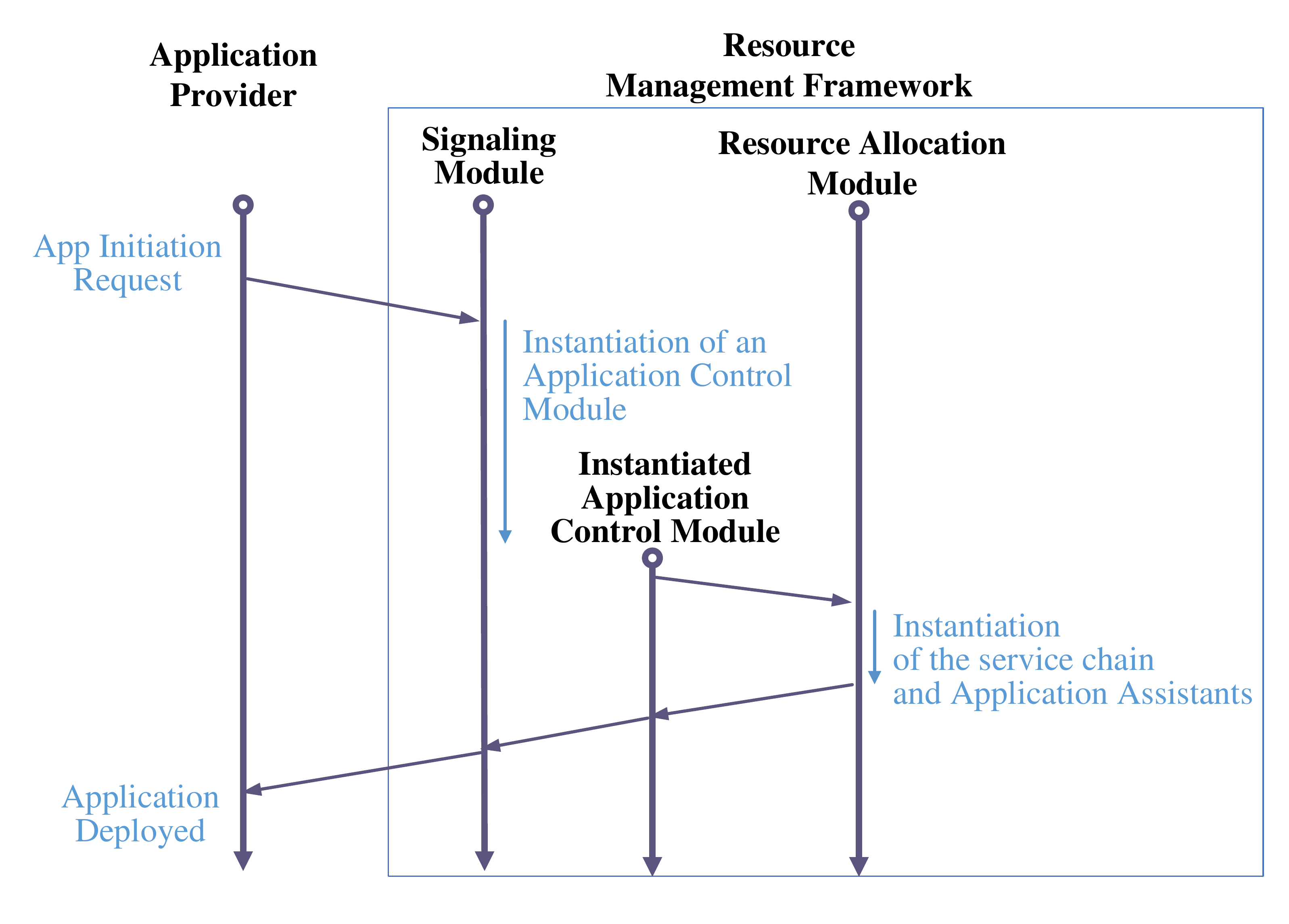}
	\caption{Signaling - application instantiation process}
	\label{fig:Signaling}
\end{figure}
%--------------------------------------------------------------------------------

$-$	\textsl{Signaling module:} the signaling module is responsible for instantiating the application. This includes the instantiation of~its~associated service function chain and a special module in the management framework called the application controller. As shown in~Fig.~\ref{fig:Signaling}, the application's provider/owner sends a request to the signaling module, which creates an instance of~the~application control module. The application control module communicates then with the resource allocation module that~will~allocate the resources for the chain.
%request the instantiation, within the network, of the application assistants and the service chain associated to the application. In the following, we provide more details about these components.

$-$ \textsl{Application Control module:} this module is part of the management framework. 
An instance of the application control module is created for each application (Fig.~\ref{fig:ManagementFramework}).
This module is in charge of managing the~SFC that processes and~delivers~the~traffic of~a~particular application.
It is hence~evaluates the traffic demand and~the~amount of~resources required for the chain. 
This~means that~it~responsible for~estimating the number of instances (virtual machines or containers) for each network function in the chain and~their~resource requirements in terms of CPU, memory and disk and the amount of bandwidth required for each virtual link. 
To achieve this goal, the application control module regularly receives monitoring data from the monitoring module and also additional metadata (for instance, information describing the~run-time application's requirements) and~commands from the application assistants (located at the application endpoints). Based on these information, the Application Control module could scale up and down the resources of network functions, add more network functions in the chain and might configure the network function operation (e.g.,~changing the forwarding rules for a particular routing network function, changing the~encoding scheme in~a~video-encoding network function).

%This module tightly collaborates with application transport assistants of each application (Fig.~\ref{fig:OSAP}) in~order to ensure the requested performance. In this context, Machine Learning techniques could be used to dynamically predict applications' requirements based on historical traces and to estimate the amount of resources needed to process the traffic. In addition, dynamic resource optimization schemes could be proposed to leverage virtual machine migration to dynamically relocate virtual machines or containers in order to achieve the sought-after objectives (e.g.,~performance, cost).

$-$ \textsl{Resource allocation module:} this module is responsible for allocating resources requested by the application control module while ensuring the network operator's high-level objectives (e.g.,~performance, network utilization,  energy efficiency and operational costs).
This module could also implement resource optimization schemes to leverage virtual machine migration to~dynamically relocate virtual machines or containers in order to achieve the sought-after operator's objectives.

$-$ \textsl{Failure management module:} this module is responsible for identifying and predicting failures and mitigating their impact on~the~deployed services in a proactive or reactive manner. It should take into account failure patterns and the availability requirement for each service chain in terms of~number of~nines.

%deployed over the infrastructure. 

$-$ \textsl{Monitoring module:} this module is in charge of collecting  different statistics about the physical and virtual components and monitor their actual state. The type of the~gathered data and the collection frequency could be configured and adjusted by~the~application control module and the resource allocation module to collect statistics for on the service function chains and~the physical infrastructure.

%The design of these modules is a primary challenge that should be solved during the project. Of course, the proposed solutions should be scalable and able to ensure application performance requirements while achieving several objectives related to performance, service level agreements, scalability, and revenue.

%-------------------------------------------------------------------------------------
\subsection{Business Model}\label{sec:BusinessModel}
%----------------------------------------------------------------------------------------
Future Internet will involve different stakeholders. Although they might look similar to the ones of today's Internet, their~roles, responsibilities and the services they are offering will significantly evolve. We can identify the following two main stakeholders:

$\bullet$ \textbf{Network operator}: similar to an ISP, a~network operator owns and~manages the~physical infrastructure and the virtual resources. However, the~services offered by such operators should not be restricted to data delivery but it should encompass more services and stringent guarantees. These services could be service function chains or network functions that might be connected in~a~particular topology with stringent requirements in terms of performance  (e.g.,~cpu, memory, disk, end-to-end delays, processing delays, rate, packet loss), security, reliability and availability.
The network operator is supposed to deploy platforms and software required to run the requested network functions and to provision the~requested service function chains. 

$\bullet$ \textbf{Application provider}: an application provider is a~company or~an~institution that want to deploy a particular application and~offer it to end users. He~defines the required service function chain adapted to the application to~be~deployed. Hence, he~is~in~charge of~identifying the chain components, topology and potential sources and~destinations of the traffic. An~application provider relies on~a~network operator to provision and manage the requested chain and to satisfy the~potential service level agreement. In some cases, the application provider could be a~network operator at the same time.

$\bullet$ \textbf{End user}: end users are users of a particular application. The traffic originated from these users will be steered across the service function chain deployed for the application.

%\begin{figure}[!hbt]
	%\centering
		%\includegraphics[width=0.60\textwidth]{Figures/InternetFuture.pdf}
	%\caption{Example of an application deployment over several networks}
	%\label{fig:InternetFuture}
%\end{figure}

%Fig.~\ref{fig:InternetFuture} shows an example where an application is deployed over the infrastructures of several network operators as multiple service function chains connected to each other. 

%=========================================================
\section{Use cases}\label{sec:UseCases} 
%=========================================================

In this Section, we provide some use-case applications where we showcase how FlexNGIA could implement them and satisfy their requirements.

%----------------------------------------------------------------------------------------
\subsection{Mixed Virtual Reality and Holograms} \label{sec:MixedVirtualRealityWithHolograms}
%----------------------------------------------------------------------------------------

%---------------------------------------------------------------------
\begin{figure}[!htbp]
	\centering
		\includegraphics[width=0.85\textwidth]{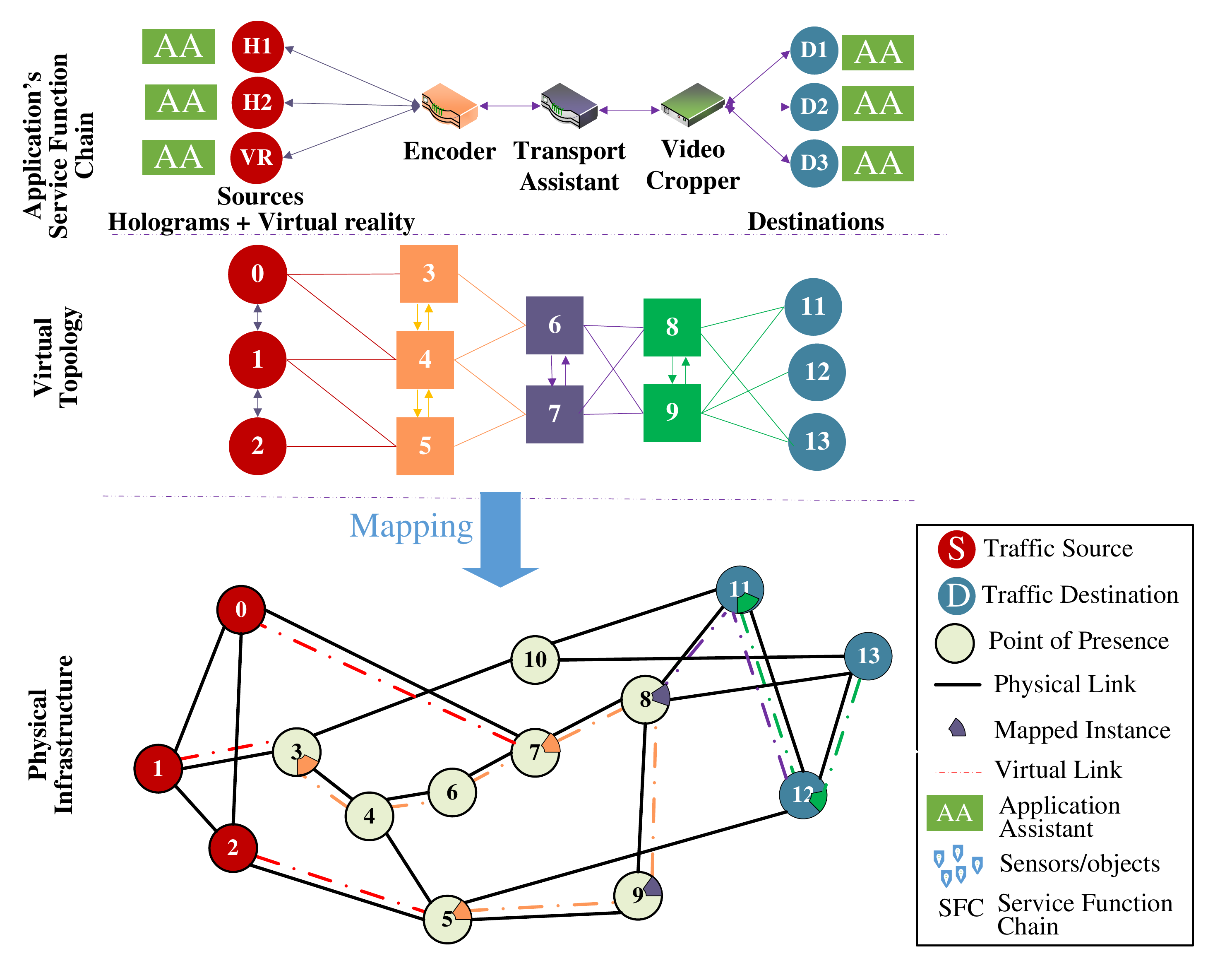}
	\caption{Mixed reality with virtual reality and holograms}
	\label{fig:MixedVRHologram}
\end{figure}
%---------------------------------------------------------------------

In this use-case, we are considering a virtual reality environment in which several human holograms should be incorporated (Fig.~\ref{fig:MixedVRHologram}). The whole virtual environment should be ``ported'' at real-time to three users situated in different geographical locations. 
The users should be then able to explore the virtual environment and interact with the teleported human holograms.  

Fig.~\ref{fig:MixedVRHologram}~shows the service function chain required for this application. The chain has three sources of data (two~holograms H1 and H2 and one virtual reality environment VR1) and three destinations where users are located at~D1, D2 and~D3. In~this~example, we~assume three network functions are needed, namely an encoder to encode and compress the~3D~video/audio flows, a transport assistant, and a~video cropper. In practice, there might be several instances of~each network function. The~example shown in Fig.~\ref{fig:MixedVRHologram} assumes that there are three instances of~the~encoder, two instances of~the~transport assistant and two video croppers. These instances are then mapped onto the physical infrastructure taking into consideration several potential objectives like the~performance (e.g.,~end-to-end delay), energy efficiency and availability.

In the following, we provide more details about these three network functions and how they should operate for this particular application:

$\bullet$ Encoder: this network function is charge of encoding and compressing the 3D video/audio flows coming from the three sources. In the example provided~in~Fig.~\ref{fig:MixedVRHologram}, we assume there is an instance of the encoder for each source of data. 

$\bullet$ Transport Assistant: this function is in charge of the congestion control, packet caching and retransmission, packet scheduling and dropping decisions. We can see in this example that two instances of the Transport Assistant are created in physical nodes 8 and 9, respectively. For example, as~the instance in node 8 receives packets coming from the three sources. It is aware of all the flows coming from the three sources and can prioritize or drop packets based on~the~information provided by the application assistants. Furthermore, Transport Assistant instances could also cache packets requiring low latency and retransmit them directly in case of~packet loss. Each of these instances could also implement a congestion control and retransmission strategy depending on the application's requirements.

$\bullet$ Video Cropper: this network function operates at the application layer and is responsible for cropping some parts of the virtual environment or holograms when needed and could adjust the video/audio quality through compression. This allows to reduce the size of the transmitted data and only keep the most relevant data needed by a particular user. For instance, when user~1 is interacting with hologram H1 only and is looking at~a~particular angle of it (e.g.,~face), the video cropper can crop the remaining angles or parts of the hologram. At the same time Hologram H2 maybe interacting with~user~2. In this case, the cropper cuts down or reduces the quality of the other parts of the virtual environment before sending the data to~user~2. As~a~result, the~benefit of~the~cropper is to reduce the amount of the~data transmitted to a user taking into account the current context of~the~application based on the interest and behavior of the user.

It is also worth noting that identifying the number of instances, their order and their placement is an open research challenge that should be~addressed while taking into account the application's context and requirements in terms of~performance, reliability and availability.

%----------------------------------------------------------------------------------------
\subsection{Network-Assisted Reliable Data Transport} \label{sec:NetworkAssistedDataTransport}
%----------------------------------------------------------------------------------------

%--------------------------------------------------------------------------------
\begin{figure}[htbp]
	\centering
		\includegraphics[width=0.95\textwidth]{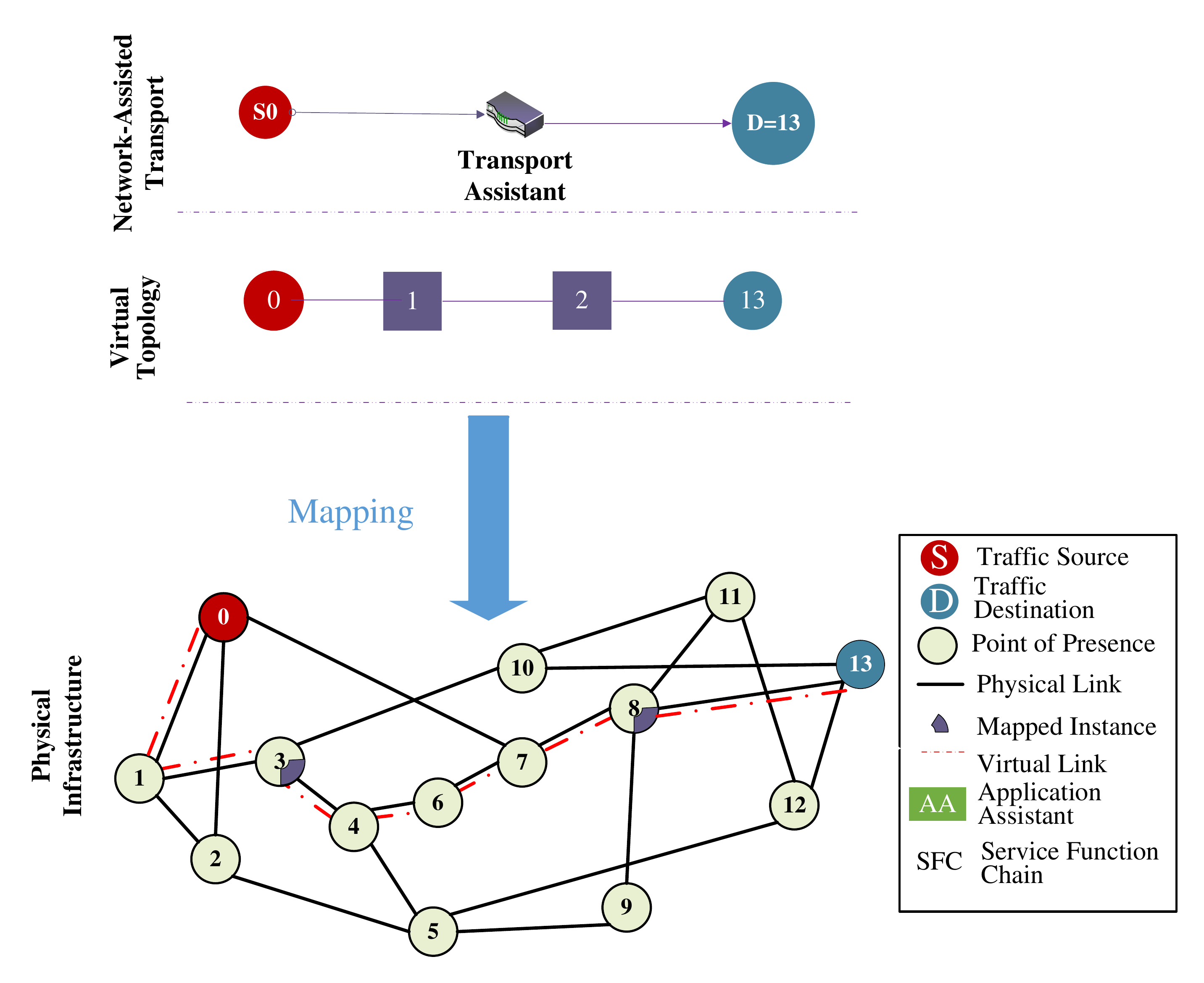}
	\caption{Network-Assisted data transport }
	\label{fig:NATCP}
\end{figure}
%--------------------------------------------------------------------------------

In traditional networks, TCP ensures reliability through retransmissions when packet loss is detected. However, current TCP is not really able to accurately detect packet loss and only speculates about it when a timeout is reached or when three~duplicate acknowledgments are received. This may incur more traffic (e.g.,~unnecessary retransmissions) and additional delays for retransmitting lost packets, leading for more congestion and higher latency in the network. 

The combination of layers 3 and 4 allows the combined layer to be aware of the state of the network (congestion) and~the~current loss of~packets and~therefore, it could better manages congestion. For instance, the network could ensure the~reliability instead of TCP. In this case, many instances of the \textit{Transport Assistant} could be created in the path towards the~destination to offer reliability by detecting packet loss and carrying out retransmission from the network nodes (Fig.~\ref{fig:NATCP}). A Transport Assistant could be seen as a new network function that is responsible for~assisting the transport of data by providing other functions like congestion control and reliability.
One or more instances of the Transport Assistant could be instantiated in the path between the source and destination depending on the application's requirements (e.g.,~in terms of performance and reliability).
More precisely, here are some services that could be offered by~this~function:

$\bullet$	The Transport Assistant could keep a copy of the transmitted packets (i.e.,~packet caching) and analyze incoming traffic to check for acknowledgments. If a packet loss is detected, either by an explicit message or through a timeout or duplicate acknowledgments, the Transport Assistant retransmits the lost packet without the need to wait for the source to retransmit~it. This significantly reduces retransmission delays and allows to detect more accurately packet loss. It is worth noting that this~new function acts at the transport level and breaks the traditional end-to-end principle of the traditional OSI transport layer. Of~course, more sophisticated packet retransmission mechanisms could be designed and implemented in~Transport Assistants.

$\bullet$ Any instance of the Transport Assistant throughout the path could also implement a congestion control scheme that adjusts the rate at which packets are forwarded depending on the state of the remaining portion of the path (i.e.,~congestion). This~feature is different from traditional TCP where only the source node is responsible for adjusting the data sending rate.

$\bullet$ The header of the packets could be customized to provide the transport assistants with the required information to carry out the aforementioned services. For instance, the header could contain metadata that provide information about the different video layers or the components of a hologram (e.g.,~layer or component importance and priority, requested reliability for~each packet, interval of time for which the packet is needed). This will help the Transport Assistant to take informed decisions about the~incoming packets (e.g.,~dropping packets of some layers or components if needed, deciding whether a~packet should retransmitted or not).
%or compressing video
%video encoder type, 

%$\bullet$ The transport assistant is aware of the application requirements. For instance, if the data transmitted is video, it could apply compression techniques to further reduce the content size if there is a congestion in the remaining path to the destination. It could also selectively reject packets depending on their content without the need to retransmit them. 

%----------------------------------------------------------------------------------------
\subsection{Deterministic Networking} \label{sec:DeterministicNetworking}
%----------------------------------------------------------------------------------------

In this use case, we focus on the deployment of a Deterministic Networking (DetNet) service over the FlexNGIA architecture. DetNet services \cite{ietf-detnet-architecture-13} have been proposed by IETF DetNet Working Group with the goal of carrying unicast and multicast data flows with extremely low latency and low data loss rates within a single network domain. In DetNet, Quality of~Service (QoS) parameters are defined in terms of the minimum and maximum end-to-end latency and jitter, packet loss ratio, and an upper bound on~out-of-order packet delivery. 
%The DetNet focuses on worst-case values for the end-to-end latency, jitter, and misordering.

DetNet ensures the QoS objectives thanks to three techniques, namely, (1) resource allocation, (2) explicit routing and (3) protection mechanisms. In the following, we provide more details about these techniques. 
Resource allocation refers to allocating enough bandwidth and buffer to each flow. Explicit routing means that the route of~the~data flow does not change during the flow lifetime.  Protection service aims at controlling packet loss by mitigating media and memory errors and failures. It mainly relies on the key idea that loss can be significantly reduced by sending the data over multiple disjoint paths. This is achieved by a set of network functions that are described in Table~\ref{tab:DetNetNF}. 

%They~include the Packet Replication Function~(PRF) that allows to send multiple packet duplicates through different paths, the Packet Elimination Function~(PEF) that in charge of eliminating duplicate packets, and the Packet Ordering Function~(POF) that uses the sequencing information to re-order a flow's packets that are received out of order. The Packet Encoding Function~(PEF) that encodes the information of~a~packet within multiple packets,

%-------------------------------------------------
\begin{table}[!t]
\centering
\caption{DetNet Network Functions}
\label{tab:DetNetNF}
\begin{tabular}{|l|l|}
\hline
\textbf{Function Name}      & \textbf{Function Operation} \\ \hline
Packet Replication Function (PRF) & \begin{tabular}[c]{@{}l@{}}Send multiple packet duplicates\\through disjoint paths\end{tabular}    \\ \hline
Packet Elimination Function (PEF)      & Eliminate duplicate packets \\ \hline
Packet Ordering Function (POF)    & \begin{tabular}[c]{@{}l@{}}Re-order packets that are received out of order\end{tabular} \\ \hline
%Packet Encoding Function  (PEnF) & \begin{tabular}[c]{@{}l@{}}Encode the information of one packet within multiple packets in different member flows\end{tabular} \\ \hline
%Packet Decoding Function  (PDF) & \begin{tabular}[c]{@{}l@{}} Take packets from different member flows, and computes the original packet\end{tabular} \\ \hline
\end{tabular}
\end{table}

%-------------------------------------------------
%---------------------------------------------------------------------
\begin{figure}[!b]
	\centering
		\includegraphics[width=0.90\textwidth]{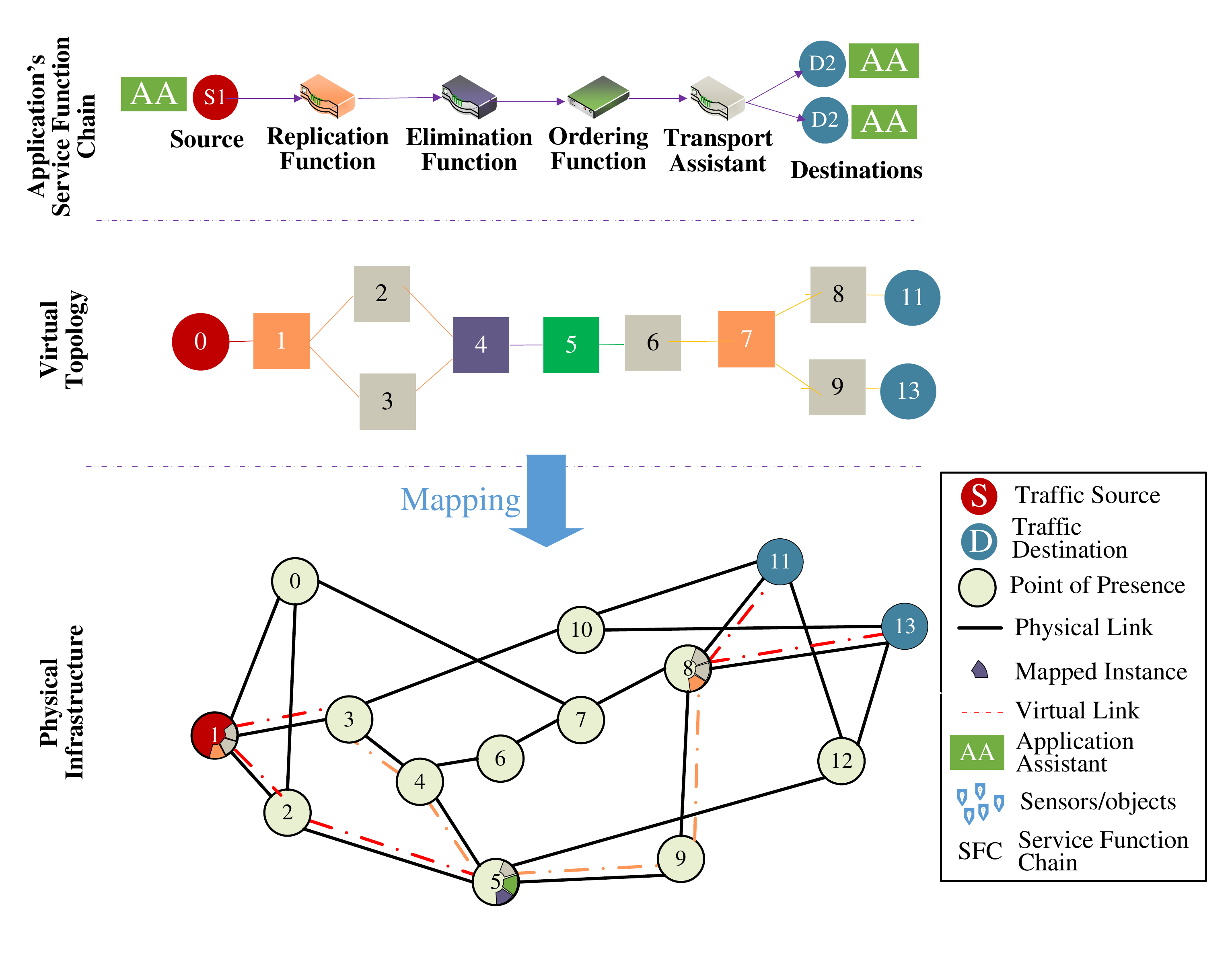}
	\caption{Example of Deployment of a Deterministic Networking Service}
	\label{fig:DetNet}
\end{figure}
%---------------------------------------------------------------------
Figure~\ref{fig:DetNet} shows a typical DetNet Service with its typical network functions (the ones described in~Table~\ref{tab:DetNetNF}) and empowered by the Transport Assistant function. The traffic should be delivered from a source (POP 1) to two destinations (POPs 11 and~13) with a pre-determined end-to-end latency and jitter, close-to-zero packet loss ratio, and an upper bound on~out-of-order packet delivery.
The figure shows how the service function chain is translated into a virtual topology, which is, in turn, mapped onto~the~physical infrastructure. We can see that POP 1 hosts a replication function that replicates the traffic into two different paths to reach POP 5 where an elimination function ensures to keep a single copy of each packet by dropping duplicated packets. The same POP 5 hosts a~packet ordering function  that reorders the packets before sending them to POP 8 to~be~duplicated towards the~destinations (POPs 11 and~13).

To further improve the service performance and reliability, we included the Transport Assistant function in~the~service function chain. Thus, several instances of the Transport Assistant (placed in~POPs~1,~5 and~8) will carry out~services of~the~combined transport and network (e.g.,~packet caching and retransmission, and~congestion control).

It is worth noting that the number of instances for each function as well as the order of the functions are provided as~an~illustration in~this~example. More research work should be carried out to investigate the best techniques to identify such parameters and evaluate their impact on the service performance. 
%----------------------------------------------------------------------------------------
\subsection{High-Precision Monitoring and Packet Scheduling} \label{sec:HighPrecisionMonitoring}
%----------------------------------------------------------------------------------------
In this use-case, we focus on high precision debugging and monitoring. We also look at the possible options to leverage FlexNGIA to efficiently collect statistics in-network and leverage high precision monitoring to provide high precision packet scheduling. In the following, we provide more details on how these functionalities could be implemented:

\textbf{$\bullet$ In-network and high-precision statistics collection:} network functions could be configured dynamically to incorporate high precision statistics into data packets (e.g.,~timestamp, processing  and queuing delays, queue size). %These statistics could also be collected by a monitoring function included in~the~chain. 
Another possibility is to include a monitoring function in the service function chain that could collect such statistics, incorporate them into~the~packets and eventually report them to the application control module or other functions in the chain. This function could also analyze the statistics and report the analysis outcome only when it is needed (e.g.,~based on a threshold on some statistics related to~the~network or~the~application performance). This could significantly reduce the monitoring traffic and the need to~communicate with the application control module while it reduces the time needed to report potential issues. 

%\textbf{$\bullet$ High-precision debugging:} network functions or specific monitoring functions could be instantiated to add the needed commands to ask the following network functions to include metadata with high-precision statistics (e.g.,~processing  and queuing delays, queue size) into all packets or a subset of them. This will allow to quickly analyze.

\textbf{$\bullet$ High-precision packet scheduling:} high-precision monitoring and collection of~statistics that are~incorporated into~packets (e.g., timestamp, queuing delay) could provide a more accurate and precise information to packet schedulers and congestion control mechanisms. This opens the door for~many~research opportunities to design better packet scheduling techniques in~order to~offer high-precision packet delivery (i.e.,~deterministic deadline or within an interval of time). 
Network functions could be also designed to react taking into~account these statistics. For instance, the~time remaining for~a~packet before reaching the~deadline could be used to take decisions on how to adjust the packets' queuing priority, change its route, or compress its~content.%or~combine it with another packet.

%(or such statistics  a command into the packet asking next functions to forward statistics report to another monitoring function in the chain or to the application control module, which will react depending on the received statistics to update the priority.

%----------------------------------------------------------------------------------------
%\subsection{Supporting Legacy Internet} \label{sec:SupportingLegacyInternet}
%----------------------------------------------------------------------------------------

%---------------------------------------------------------------------
\section{Key Research Challenges} \label{sec:KeyResearchChallenges}
%---------------------------------------------------------------------

% in network computing: advanced network functions
% application-aware networks : flows (multiple src multiple dest) + flow requirements 
% application communicate with the network (VNF and res management)-> this is what i need
% semi-distributed contorl and management planes 
% header -> format free headers
In this Section, we summarize key research challenges that should be addressed in order to further develop FlexNGIA architecture and ensure the performance of~the~resource management framework and the sought-after application objectives in~terms of~performance, reliability and availability.

%----------------------------------------------------------------------------------
\subsection{Designing Service Function Chains} \label{ssec:SFCDesignTrafficRouting}
%----------------------------------------------------------------------------------
One of the main challenges pertaining to the FlexNGIA architecture is how to design the service function chains that~are~tailored to~the~applications. %In other words, it is . There is no a single solution that fits all application. Hence 
Depending on the application type and requirements, the application designer should identify the best type of functions, the virtual links to connect them, and define how they should cooperate with each other and what kind of~data, metadata and commands they should exchange (e.g.,~application data, statistics, synchronization data). It is worth noting that these functions could operate at different layers (e.g.,~network, transport or application), which provides a wide range of~flexibility but makes the design of the chain more challenging. The use cases presented in~Section~\ref{sec:UseCases} show some examples of how service chains should be tailored and customized to~applications.
The design of the chain encompasses also the definition of the role of the application controller including the management and control tasks that it should carry out. Such tasks could include the elasticity management, routing within the chain, and the operation of network functions.

The FlexNGIA architecture allows the application designer to customize the chain and define the best routing strategy adapted to the chain. The routing could be managed like in SDN where a single entity (i.e,~a~single controller) programs the~forwarding rules and~the operation of~the~network functions belonging to~the~chain. FlexNGIA easily supports other routing strategies like Preferred Path Routing \cite{UmaGlobecom2018} or Segment Routing \cite{rfc8402SegmentRouting} where each node (or network function through the chain) steers the~packet through a list of instructions (called segments).

%----------------------------------------------------------------------------------
\subsection{Resource Allocation} \label{ssec:SFCProvResAlloc}
%----------------------------------------------------------------------------------
From the network operator's perspective, the placement and allocation of the  resources of the applications' service function chains are key challenges to achieve high-level objectives like ensuring high availability of the resources, minimizing energy consumption, increasing the usage of green energy and minimizing different operational costs. A large body of work has addressed some of~these challenges in the context of virtual network embedding problem and service function chaining \cite{Amokrane2013Greenhead,HarmonyJournal2013,BariCST13,chowdhury2010survey,Aidi-CNSM2018,mijumbi2015design}. However, more research effort should be performed to estimate the amount of the resources needed for~each network function in the chain, to decide  whether these functions should be distributed or not (e.g.,~single or~multiple virtual machines/containers) and to~identify the best type of hosting environments (containers, virtual machines, dedicated hardware). Of course, several parameters need to be taken into account such~as the performance requirements (e.g.,~processing time, end-to-end delays) and~also~application-level requirements (e.g.,~video quality, reliability). 

Furthermore, as mentioned previously, FlexNGIA assumes resource allocation is carried out in two steps, chain translation into a virtual topology and then the mapping of this virtual network. Though each of these steps could be carried out independently from the other, it would be interesting to explore solutions that could perform  these steps in a coordinated manner. This~is~a~promising research avenue as the translation of the chain into a virtual topology (i.e., VM/container instances and virtual links)  should ideally take into consideration the available capacity and the type of resources within the physical infrastructure.% and hence, the potential mapping of the resources. 

% Requirement evaluation and resource allocation
% mapping inherently provides potential routes
%It is all about the application use-case: looking into the functions, the input and output traffic, gain in time computing/ transmission time, imagining what kind of function could further improve the service. the most prminent in FlexNGIA is the transport assistant (multiple transport assistant synchronizing between them) AI 
%----------------------------------------------------------------------------------
\subsection{Signaling} \label{ssec:Signaling}
%----------------------------------------------------------------------------------

Signaling refers to the process of sending control packets in order to initiate the resource allocation process for a particular application and inform the network and the management framework about the main features and requirements of the application.
As described in the FlexNGIA network management framework, when a new application is instantiated, the application controller should be instantiated and the needed amount of the computing and networking resources should be allocated in the network infrastructure. It is therefore necessary to put forward efficient signaling scheme that will efficiently inform the network (i.e.,~central resource management) and eventually configure the network. The main challenge in this context is to put forward signaling schemes that allow the allocation of the service chain resources and performance requirements of~the~application that should be ultra-fast and efficient.

The signaling becomes more challenging when resources need to be allocated throughout different networks belonging to~different network operators \cite{tusa2018multi,galis2018network}. 
In this case, more advanced signaling schemes could be proposed to include automated negotiation process between different network operators to identify the ones that should be involved to allocate the resources for a particular request and to identify the best end-to-end resource allocation in terms of performance, reliability and security guarantees and~in~terms~of~the~operational~costs~as~well.

%----------------------------------------------------------------------------------
\subsection{Distributed Cross-Layer and Advanced Network/Transport Protocols} \label{ssec:DistributedTransportProtocol}
%----------------------------------------------------------------------------------

The Transport Assistant is a network function introduced by FlexNGIA to allow the network to offer combined transport and~network services, which will allow faster packet retransmission through packet caching and better control over the~congestion as it is managed from within the network.  

In the use-case of network assisted reliable data transport (Section \ref{sec:NetworkAssistedDataTransport}), we discussed how Transport Assistants proposed in~FlexNGIA could achieve such goal. However, there is still several challenges that should be addressed to fully benefit of~the~cross-layer transport. For instance, one of these challenges is to identify the optimal number of~Transport Assistant and their placement within the service chain. It is also of utmost importance to define a distributed cross-layer protocol to~ensure the~communication between the Transport Assistant instances that will allow them to interact with each other and exchange information about the state of the communication and the progress of data delivery (e.g.,~acknowledgment, data sequence, packet loss, congestion control) and~also devise novel and customized congestion control, packet caching and retransmission algorithms that are tailored to the application's requirements.

%\textcolor{blue}{
Beyond the possibility of designing distributed cross-layer protocols to leverage the Transport Assistants, the FlexNGIA architecture allows to deploy any protocol operating on top of layer 2 including network and transport layers and to fully benefit from the softwarization technologies to easily adopt existing protocols (e.g., Multipath TCP), to evolve them and also to devise new ones that are tailored to the application's requirements.
\subsection{Fault-tolerance and Failure management} \label{ssec:FailureManagement}
%----------------------------------------------------------------------------------
%Application controller, cooperating network functions 
To ensure high availability of the future applications and services, several failure management solutions should be developed. They~can~be~either proactive or reactive \cite{RabbaniIEICE13,Aidi-CNSM2018,ZhangVenice2014,Ayoubi2016,SVNE2013,Zhani2015Surv}. Proactive techniques aim at addressing failures before they happen. This can be done through provisioning backup virtual machines and backup links that can take over the service in~case~of~failure. The main benefit of such techniques is that service interruption time is minimal; however, this comes at~the~cost of allocating backup resources that are only used when a~failure~happens. Unlike proactive techniques, reactive techniques aim at mitigating failures after they happen. Of course, these techniques incur less resource wastage but may result in a higher interruption time.

%\textcolor{blue}{
Furthermore, more work is needed to achieve the high availability of network functions and to put forward efficient recovery schemes that allow a fast, smooth and automatic failover after a failure and ensure a quick data recovery and state reconstruction in the backup function \cite{Sherry2015,Zhani2015Surv}.
%}

%----------------------------------------------------------------------------------
\subsection{High-Performance Virtual Network Functions and Services} \label{ssec:HighPerfVNF}
%----------------------------------------------------------------------------------
In order to cater to the stringent performance requirements of future applications, network service chains processing the~exchanged packets need to consist of ultra-low latency software or hardware network functions. In this context, software-based network functions running on virtual machines and containers hosted in commodity servers seem to be the best option to~provide high flexibility to create network functions whenever and~wherever needed. 
%\textcolor{blue}
However, the performance of such network functions in terms of instantiation time, processing capacity and latency is still low compared to hardware and still hard to be predict. This precludes their deployment in production environments and brings the need to further investigate server and NIC virtualization technologies, Linux kernel network stack,  I/O and~CPU schedulers, and inter-VM communication in order to maximize the packet processing capacity and~speed of~the~containers and~virtual machines and to ensure high isolation among them in terms of CPU and networking performance~\cite{Zhang2016OpenNetVMAP,hwang2015netvm,liu2017design,ToussainHipNet2018,HongHipNet2018,Kumar2019}.

Furthermore, network functions could be also implemented in dedicated hardware and programmable data plane chips. In~this~context, a challenging task is to dynamically design and develop different kind of network functions in an optimal and~efficient manner on programmable hardware using available languages like P4~\cite{BosshartP42014}.
 
%We need to look into dynamically programming the hardware to add the new functions that are needed by the application. 

%P4 is a high-level language for programming protocol-independent packet processors
%This inter-VM communication

%able to process the data (P4, NPU,)
%. OpenNetVM runs network functions in lightweight Docker containers, enabling fast startup and reducing memory overheads. r high-speed inter-VM communication through shared huge pages and enhancing the CPU scheduler to prevent overheads caused by inter-core communication and context switching
%Moving packet data involves delivering the packet from the network interface to an NF, moving it across functions on the same host, and finally across yet another network to NFs running on other hosts in a cluster/data center. 

%----------------------------------------------------------------------------------
%\subsection{High-Precision Time Synchronization} \label{ssec:HighPrecisionTelemetry}
%----------------------------------------------------------------------------------
%It is also worth noting is that high-precision time synchronization is one of the major challenges pertaining to monitoring and also as other management tasks.  

%----------------------------------------------------------------------------------
\subsection{High-Precision Monitoring and Measurements} \label{ssec:HighPrecisionTelemetry}
%----------------------------------------------------------------------------------
Monitoring and measurements are key features not only to debug the potential issues with the infrastructure and the applications but also to provide the customers of the network operators with the possibility to check that the SLA requirements associated with~the~requested service chain are respected 
(in~terms of delay, bandwidth and packet loss).
In addition to traditional coarse grained monitoring of high-level network services and traffic flows, future applications may require high-precision monitoring and measurement (e.g.,~at the scale of micro or milliseconds) at~a~fine-grained level (e.g.,~at~the~packet level). A~large body of work has looked at the monitoring challenge in software-defined networks and how to minimize different monitoring costs (e.g.,~number of collected packets, bandwidth, storage, processing) while reducing data collection and processing time~\cite{Chowdhury2014Payless,Shu2016,kim2015inband}. However, fine-grained and~high-precision monitoring and~measurements techniques have not been well addressed in the literature. 
Inband network telemetry should also be developed to allow data and statistics collection through the~data plane without involving the control plane \cite{InBandTel2016}.

Furthermore, to be able to offer ultra-low latency services and allow high-precision monitoring, one of the major challenges is to ensure high-precision time synchronization between the infrastructure's nodes. The synchronization error should be in~the~order of~microseconds or less. 
This will provide the ability to the network operators and to their customers as~well to~carry out high-precision and fine-grained debugging and monitoring and also to check and validate the~service level agreements in~terms~of~delay, bandwidth and~packet loss.

%----------------------------------------------------------------------------------
\subsection{SDN++} \label{ssec:SDNplusplus}
%----------------------------------------------------------------------------------
%APIs P4 and more...
The concept of software defined networking has been evolving since it is emergence. 
First, the OpenFlow protocol \cite{McKeown_OpenFlow} allowed to process packets differently thanks to match-action rules ``if the packet matches a condition, then do this action''. Data plane programmability using P4~\cite{BosshartP42014} or Protocol Oblivious Forwarding (POF)~\cite{Haoyu2013,Li2017}~goes further by providing an entire language to instruct the networking equipment how to process packets. For~instance, it~could tell the switch or router how to parse the packet, which fields to match, how to encapsulate, and what action to perform for each header. This opens up the~possibility that the entire network could be reshaped within a single administrative domain.

FlexNGIA advocates to further broaden the concept of software defining networking by providing protocols and languages that allow to configure and program the behavior of network functions (referred to~as~SDN++). Advanced languages could be developed to allow to program the services offered by the transport assistants (e.g.,~caching, retransmission algorithms, congestion control) or by other application-level functions (e.g.,~hologram or video croppers, data compression, data~aggregation). 

The development of SDN++ will also push forward the concept of intent-driven networking \cite{ElkhatibIntentCNSM2017,clemm-nmrg-dist-intent-01} where  high level requirements (called intent) are conveyed to the network. In this context, the major challenge is how to express an intent and~how to~translate it into a service function chain, network functions, routing schemes, and protocols that would allow to ensure the intent's requirements.

%Intent — the idea of policy-driven networking — is getting a lot more of the spotlight than OpenFlow these days, but it looks like P4 can fit into that world.

%This allows the switch or router to recongnize
%P4 lets us define what headers a switch will recognize (or "parse"), how to match on each header, and what actions we would like the switch to perform on each header. For example, we might tell the switch to process  IPv4 headers by 
%P4 broadens that idea by providing an entire language for telling the switch or router what to do. Rather than fill out tables, you would write a P4 program to control the equipment. The program could include add-ons such as analytics.

%Unlike OpenFlow, P4 is not a protocol but rather a programming language that allows to program the data plane. That is to instruct the networking equipment what to do with packet (e.g, of packet forwarding planes

%----------------------------------------------------------------------------------
\subsection{Pricing} \label{ssec:Pricing}
%----------------------------------------------------------------------------------
As FlexNGIA advocates offering service function chains to steer the traffic of each application, pricing becomes more challenging than it is for today's Internet that offers only data delivery as a service.
%Current \cite{Ghrada-PVESDN2018}.
Novel pricing scheme should be devised to take into account not only the amount of traffic steered through the chain but also the amount of computing and networking resources allocated for the chain as well as their usage over time. It is also important to note that the amount of resources is directly related to the desired service level agreement in terms performance, availability and reliability of the service chain. It~is~therefore of~utmost importance to develop pricing models that estimate a fair price for a~service function chain depending on~the~type of its composing network functions, the virtual links connecting them, the amount of resources they are consuming and~the~requested SLA.

% account these parameters and to ensure a fair

%----------------------------------------------------------------------------------
\subsection{Security and Privacy} \label{ssec:SecurityAndPrivacy}
%----------------------------------------------------------------------------------

%trust?
%sharing the key
%processing data within packets?

%data delivery with stringent requirements without access to the data.
As FlexNGIA promotes in-network computing, many network functions should access the header and the content of~the~packets to be able to read, process, use and eventually modify them if needed. While this feature offers a high flexibility to devise new network functions, security, privacy and trustworthiness remain daunting challenges to address. These challenges have been widely discussed when cloud computing model and services emerged few years ago \cite{Gessner2012,Takabi2010,sun2011surveying}. However, there is still a lot of work to be done in order to make sure that the cloud  and the proposed in-network computing functions are privacy-preserving, secure and trustworthy.

%it brings the challenges that has been raised the last recent years for the cloud computing in general 
%Trustworthnisss, 
%One reseach task would be sharing keys 
%This could be solved by proposing mechanisms and solutions to efficiently share encryption keys between network functions to allow to access to the packets' content and headers.
Many solutions could be envisaged to secure the FlexNGIA architecture and the network functions of the provisioned chains. For instance, a promising solution would be to define levels of security to identify the different access levels to~the~packet header and content. In this case, some network functions may not have the right to read and modify the packet header or data, others may only have access to some fields in the header (i.e.,~the~remaining header fields could be encrypted) or to particular types of the packets.
For example, a traffic management function would need only to read the priority of the packet without having access to the encrypted content. Another function may need only to access some particular flows of packets or some types of packets within each flow (e.g.,~flows with no private data).
In this context, different techniques to~share and~distribute security keys among functions and potentially different network operators could be devised to allow to access to~different types of packets or data within the packet. 
It is clear that the security and privacy techniques and solutions differ from one application to another and should be customized based on the requirements of each application.

\begin{figure}[!h]
	\centering
		\includegraphics[width=1.00\textwidth]{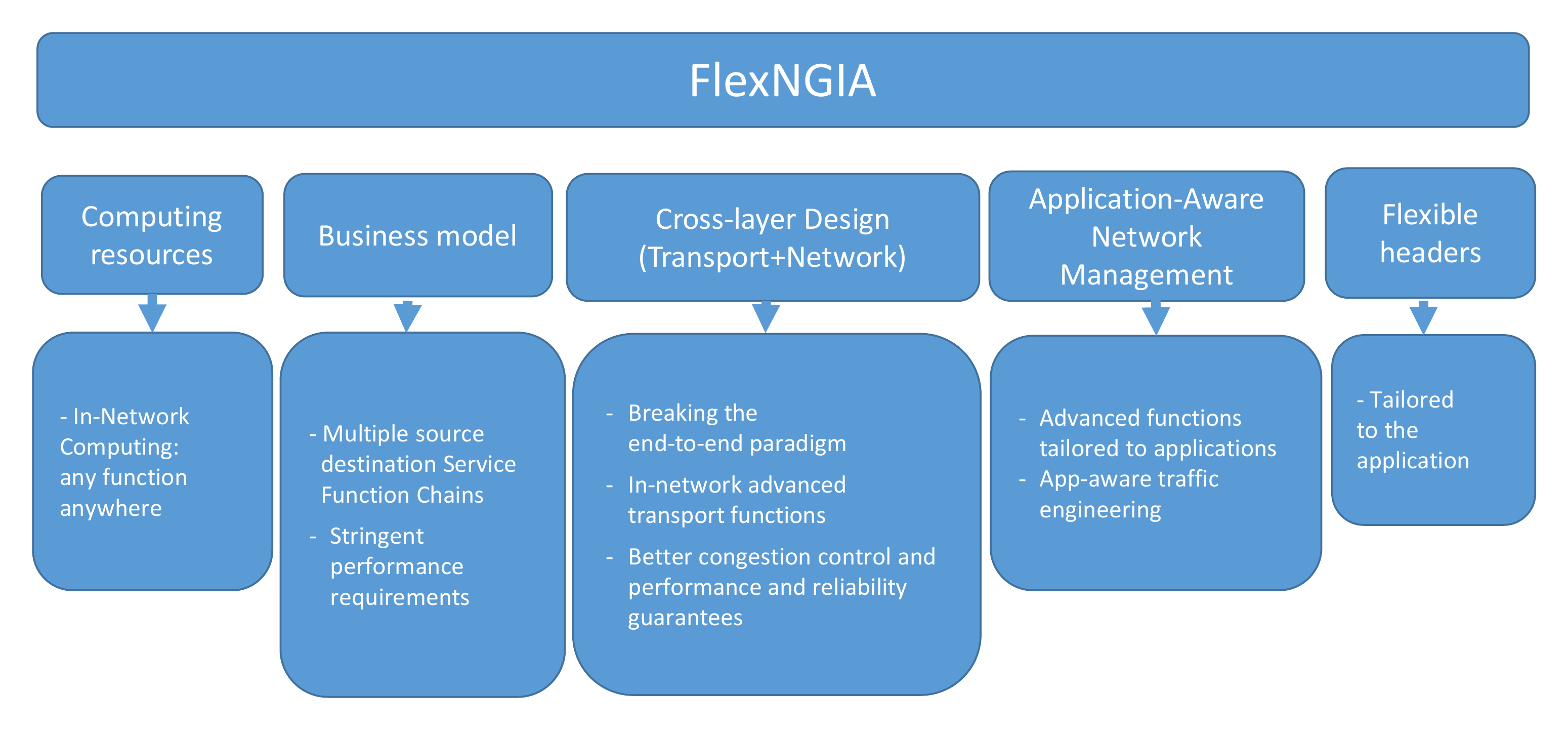}
	\caption{Main features of the FlexNGIA Internet architecture }
	\label{fig:FlexNGIA}
\end{figure}

%-----------------------------------------------
\section{Conclusion}\label{sec:Conclusion}
%--------------------------------------------------

In this paper, we analyzed the characteristics and requirements of future networking applications. It~also~highlighted the~limitations of the today's network architecture and protocols and their inability to cater to these requirements. We~hence~put forward a~Flexible Next-Generation Internet Architecture adapted to the Tactile Internet (called FlexNGIA) that could address these limitations. As~shown~in~Fig.~\ref{fig:FlexNGIA}, FlexNGIA defines a novel Internet architecture that leverages the~availability of~computing resources throughout future network infrastructures to allow in-network computing. This allows the~network to~host advanced network functions that could support the network applications. FlexNGIA also defines a business model where network operators could offer not only data delivery but also service function chains with stringent requirements in~terms of~performance, reliability and availability. Another relevant feature of FlexNGIA is that it~advocates for~the~combination of transport and network layers that allows the network to offer better congestion control and reliability services by allowing in-network advanced functions to~be~aware of the flows belonging to to same application and the requirements of~each~of~them. 
FlexNGIA considers also~a~fully flexible packet headers that could be tailored to~the~application requirements.

As discussed in the aforementioned key research challenges, FlexNGIA provides only the potential building blocks of~the~future Internet architecture that requires deterministic and ultra-low latency, high bandwidth requirements, high reliability, resiliency and~availability. However, in order to cater to these requirements, many challenges pertaining to FlexNGIA  need to be addressed including signaling, the design of service function chains, resource allocation, fault-tolerance, performance, monitoring, pricing and security.

%\begin{thebibliography}{99}
%\end{thebibliography}

%\bibliography{MyOwnRef-Journals,MyOwnRef-Conferences,References} %if bibliography
%\bibliographystyle{IEEEtran}

%\bibliographystyle{plain}

%\bibliographystyle{IEEEtran}
%{ 
%\bibliography{biblio}

\begin{thebibliography}{10}
\providecommand{\url}[1]{#1}
\csname url@samestyle\endcsname
\providecommand{\newblock}{\relax}
\providecommand{\bibinfo}[2]{#2}
\providecommand{\BIBentrySTDinterwordspacing}{\spaceskip=0pt\relax}
\providecommand{\BIBentryALTinterwordstretchfactor}{4}
\providecommand{\BIBentryALTinterwordspacing}{\spaceskip=\fontdimen2\font plus
\BIBentryALTinterwordstretchfactor\fontdimen3\font minus
  \fontdimen4\font\relax}
\providecommand{\BIBforeignlanguage}[2]{{%
\expandafter\ifx\csname l@#1\endcsname\relax
\typeout{** WARNING: IEEEtran.bst: No hyphenation pattern has been}%
\typeout{** loaded for the language `#1'. Using the pattern for}%
\typeout{** the default language instead.}%
\else
\language=\csname l@#1\endcsname
\fi
#2}}
\providecommand{\BIBdecl}{\relax}
\BIBdecl

\bibitem{ITUTactileInternet}
{ITU-T Technology Watch Report}, ``{The Tactile Internet},'' 2014.

\bibitem{McKeown_OpenFlow}
N.~McKeown, T.~Anderson, H.~Balakrishnan, G.~Parulkar, L.~Peterson, J.~Rexford,
  S.~Shenker, and J.~Turner, ``{OpenFlow: Enabling Innovation in Campus
  Networks},'' \emph{ACM Computer Communication Review}, vol.~38, no.~2, pp.
  69--74, April 2008.

\bibitem{BosshartP42014}
P.~Bosshart, D.~Daly, G.~Gibb, M.~Izzard, N.~McKeown, J.~Rexford,
  C.~Schlesinger, D.~Talayco, A.~Vahdat, G.~Varghese, and D.~Walker, ``{P4:
  Programming Protocol-independent Packet Processors},'' \emph{ACM SIGCOMM
  Comput. Commun. Rev.}, vol.~44, no.~3, Jul. 2014.

\bibitem{rfc8402SegmentRouting}
\BIBentryALTinterwordspacing
C.~Filsfils, S.~Previdi, L.~Ginsberg, B.~Decraene, S.~Litkowski, and R.~Shakir,
  ``{Segment Routing Architecture},'' RFC 8402, Jul. 2018. [Online]. Available:
  \url{https://rfc-editor.org/rfc/rfc8402.txt}
\BIBentrySTDinterwordspacing

\bibitem{Obrist:2016}
\BIBentryALTinterwordspacing
M.~Obrist, C.~Velasco, C.~T. Vi, N.~Ranasinghe, A.~Israr, A.~D. Cheok,
  C.~Spence, and P.~Gopalakrishnakone, ``Touch, taste, smell user interfaces:
  The future of multisensory hci,'' in \emph{ACM Conference Extended Abstracts
  on Human Factors in Computing Systems (CHI EA '16)}, 2016, pp. 3285--3292.
  [Online]. Available: \url{http://doi.acm.org/10.1145/2851581.2856462}
\BIBentrySTDinterwordspacing

\bibitem{16KResolution}
J.~Feltham, ``{VR needs to hit 16K to match retinal resolution},''
  \url{https://www.vrfocus.com/2015/03/abrash-vr-needs-hit-16k-match-retinal-resolution},
  Accessed 3 March 2020.

\bibitem{HologramsBW2011}
X.~Xu, Y.~Pan, P.~P. M.~Y. Lwin, and X.~Liang, ``3d holographic display and its
  data transmission requirement,'' in \emph{International Conference on
  Information Photonics and Optical Communications}, Oct 2011, pp. 1--4.

\bibitem{MicrosoftHoloportation}
``Microsoft holoportation.
  \url{https://www.microsoft.com/en-us/research/project/holoportation-3/},''
  Accessed 3 March 2020.

\bibitem{Arris2016}
C.~Cheevers, M.~Bugajski, A.~Luthra, S.~MCCarthy, P.~Moroney, and K.~Wirick,
  ``Virtual and augmented reality – how do they affect the current service
  delivery and home and network architectures,'' Arris, 2016.

\bibitem{8KResolutionBW2}
K.~Iguchi, A.~Ichigaya, Y.~Sugito, S.~Sakaida, Y.~Shishikui, N.~Hiwasa,
  H.~Sakate, and N.~Motoyama, ``{HEVC} encoder for super hi-vision,'' in
  \emph{IEEE International Conference on Consumer Electronics (ICCE)}, Jan
  2014, pp. 57--58.

\bibitem{Fettweis2014}
G.~Fettweis and S.~Alamouti, ``{5G:} personal mobile internet beyond what
  cellular did to telephony,'' \emph{IEEE Communications Magazine}, vol.~52,
  no.~2, pp. 140--145, February 2014.

\bibitem{VegaHipNet2018}
M.~T. Vega, T.~Mehmli, J.~van~der Hooft, T.~Wauters, and F.~D. Turck,
  ``{Enabling Virtual Reality for the Tactile Internet: Hurdles and
  Opportunities},'' in \emph{IEEE/IFIP/ACM International Workshop on
  High-Precision Networks Operations and Control (HiPNet)}, 2018.

\bibitem{GoogleCloudAvailability}
\emph{Google Cloud: Establishing 99.99\% Availability for Dedicated
  Interconnect.
  \url{https://cloud.google.com/interconnect/docs/tutorials/dedicated-creating-9999-availability}},
  Accessed 3 March 2020.

\bibitem{rfc2475DiffServ}
\BIBentryALTinterwordspacing
D.~L. Black, Z.~Wang, M.~A. Carlson, W.~Weiss, E.~B. Davies, and S.~L. Blake,
  ``{An Architecture for Differentiated Services},'' RFC 2475, Dec. 1998.
  [Online]. Available: \url{https://rfc-editor.org/rfc/rfc2475.txt}
\BIBentrySTDinterwordspacing

\bibitem{rfc793TCP}
\BIBentryALTinterwordspacing
``{Transmission Control Protocol (TCP)},'' RFC 793, Sep. 1981. [Online].
  Available: \url{https://rfc-editor.org/rfc/rfc793.txt}
\BIBentrySTDinterwordspacing

\bibitem{rfc768UDP}
\BIBentryALTinterwordspacing
``{User Datagram Protocol (UDP)},'' RFC 768, Aug. 1980. [Online]. Available:
  \url{https://rfc-editor.org/rfc/rfc768.txt}
\BIBentrySTDinterwordspacing

\bibitem{OSI}
``{ISO/IEC 7498-4:1989 -- Information technology -- Open Systems
  Interconnection -- Basic Reference Model: Naming and addressing},''
  International Organization for Standardization (1989-11-15), ISO Standards
  Maintenance Portal - ISO Central Secretariat, 2016.

\bibitem{SCTP-rfc4960}
\BIBentryALTinterwordspacing
``{Stream Control Transmission Protocol (SCTP)},'' RFC 4960, Sep. 2007.
  [Online]. Available: \url{https://rfc-editor.org/rfc/rfc4960.txt}
\BIBentrySTDinterwordspacing

\bibitem{SCTP-rfc3758-Reliability}
\BIBentryALTinterwordspacing
M.~A. Ramalho, M.~Tüxen, and P.~Conrad, ``{Stream Control Transmission
  Protocol (SCTP) Partial Reliability Extension},'' RFC 3758, May 2004.
  [Online]. Available: \url{https://rfc-editor.org/rfc/rfc3758.txt}
\BIBentrySTDinterwordspacing

\bibitem{QUIC-UDP}
\BIBentryALTinterwordspacing
J.~Iyengar and M.~Thomson, ``{QUIC: A UDP-Based Multiplexed and Secure
  Transport},'' Oct. 2018, work in Progress. [Online]. Available:
  \url{https://datatracker.ietf.org/doc/html/draft-ietf-quic-transport-16}
\BIBentrySTDinterwordspacing

\bibitem{QUIC-HTTP}
M.~Bishop, ``{Hypertext Transfer Protocol (HTTP) over QUIC},'' Internet
  Engineering Task Force, Oct 2018.

\bibitem{quic-protocol-02}
\BIBentryALTinterwordspacing
R.~Hamilton, J.~Iyengar, I.~Swett, and A.~Wilk, ``{QUIC: A UDP-Based Secure and
  Reliable Transport for HTTP/2},'' Jan. 2016. [Online]. Available:
  \url{https://datatracker.ietf.org/doc/html/draft-tsvwg-quic-protocol-02}
\BIBentrySTDinterwordspacing

\bibitem{quic-recovery-16}
\BIBentryALTinterwordspacing
J.~Iyengar and I.~Swett, ``{QUIC Loss Detection and Congestion Control},''
  Internet Engineering Task Force, Internet-Draft draft-ietf-quic-recovery-16,
  Oct. 2018, work in Progress. [Online]. Available:
  \url{https://datatracker.ietf.org/doc/html/draft-ietf-quic-recovery-16}
\BIBentrySTDinterwordspacing

\bibitem{joseph-quic-comparison-quic-sctp-00}
\BIBentryALTinterwordspacing
A.~Joseph, T.~Li, Z.~He, Y.~Cui, and L.~Zhang, ``{A Comparison between SCTP and
  QUIC},'' Internet Engineering Task Force, Internet-Draft
  draft-joseph-quic-comparison-quic-sctp-00, Mar. 2018, work in Progress.
  [Online]. Available:
  \url{https://datatracker.ietf.org/doc/html/draft-joseph-quic-comparison-quic-sctp-00}
\BIBentrySTDinterwordspacing

\bibitem{BariCST13}
M.~F. Bari, R.~Boutaba, R.~Esteves, L.~Z. Granville, M.~Podlesny, M.~G.
  Rabbani, Q.~Zhang, and M.~F. Zhani, ``Data center network virtualization:
  a~survey,'' \emph{IEEE Communications Surveys \& Tutorials}, vol.~15, no.~2,
  pp. 909--928, 2013.

\bibitem{Elkhatib2017Fog}
Y.~Elkhatib, B.~Porter, H.~B. Ribeiro, M.~F. Zhani, J.~Qadir, and E.~Rivière,
  ``On using micro-clouds to deliver the fog,'' \emph{IEEE Internet Computing},
  vol.~21, no.~2, pp. 8--15, Mar. 2017.

\bibitem{OLANIYAN2018}
\BIBentryALTinterwordspacing
R.~Olaniyan, O.~Fadahunsi, M.~Maheswaran, and M.~F. Zhani, ``Opportunistic edge
  computing: Concepts, opportunities and research challenges,'' \emph{Future
  Generation Computer Systems (FGCS), Elsevier}, vol.~89, pp. 633--645, 2018.
  [Online]. Available: \url{https://doi.org/10.1016/j.future.2018.07.040}
\BIBentrySTDinterwordspacing

\bibitem{GoogleInfrastructure}
\emph{Google Cloud Infrastructure [Online], available
  \url{https://cloud.withgoogle.com/infrastructure}}, Accessed 3 March 2020.

\bibitem{VLAN}
``{IEEE Standard for Local and metropolitan area networks--Bridges and Bridged
  Networks},'' IEEE Std 802.1Q-2014 (Revision of IEEE Std 802.1Q-2011), pp.
  1--1832, Dec 2014.

\bibitem{rfc7348VXLAN}
\BIBentryALTinterwordspacing
M.~Mahalingam, D.~Dutt, K.~Duda, P.~Agarwal, L.~Kreeger, T.~Sridhar,
  M.~Bursell, and C.~Wright, ``{Virtual eXtensible Local Area Network (VXLAN):
  A Framework for Overlaying Virtualized Layer 2 Networks over Layer 3
  Networks},'' RFC 7348, Aug. 2014. [Online]. Available:
  \url{https://rfc-editor.org/rfc/rfc7348.txt}
\BIBentrySTDinterwordspacing

\bibitem{OTV}
\emph{{Overlay Transport Virtualization (OTV)}, \url{https://goo.gl/p8qcbj}},
  Accessed 3 March 2020.

\bibitem{Ghrada-PVESDN2018}
N.~Ghrada, M.~F. Zhani, and Y.~Elkhatib, ``{Price and Performance of
  Cloud-hosted Virtual Network Functions: Analysis and Future Challenges},'' in
  \emph{IEEE Performance Issues in Virtualized Environments and Software
  Defined Networking (PVE-SDN NetSoft 2018)}, Montreal, Canada, Jun. 25-29,
  2018.

\bibitem{ietf-detnet-architecture-13}
\BIBentryALTinterwordspacing
N.~Finn, P.~Thubert, B.~Varga, and J.~Farkas, ``{Deterministic Networking
  Architecture},'' Internet Engineering Task Force, Internet-Draft
  draft-ietf-detnet-architecture-13, May 2019, work in Progress. [Online].
  Available:
  \url{https://datatracker.ietf.org/doc/html/draft-ietf-detnet-architecture-13}
\BIBentrySTDinterwordspacing

\bibitem{UmaGlobecom2018}
U.~Chunduri, A.~Clemm, and R.~Li, ``{Preferred Path Routing - A Next Generation
  Routing Framework Beyond Segment Routing},'' in \emph{IEEE Global
  Communications Conference (GLOBECOM)}, 2018.

\bibitem{Amokrane2013Greenhead}
A.~Amokrane, M.~F. Zhani, R.~Langar, R.~Boutaba, and G.~Pujolle, ``Greenhead:
  Virtual data center embedding across distributed infrastructures,''
  \emph{IEEE Transactions on Cloud Computing (TCC)}, vol.~1, no.~1, pp. 36--49,
  January 2013.

\bibitem{HarmonyJournal2013}
Q.~Zhang, M.~F. Zhani, R.~Boutaba, and J.~L. Hellerstein, ``Dynamic
  heterogeneity-aware resource provisioning in the cloud,'' \emph{IEEE
  Transactions on Cloud Computing}, vol.~2, no.~1, pp. 14--28, January 2014.

\bibitem{chowdhury2010survey}
M.~Chowdhury and R.~Boutaba, ``{A Survey of Network Virtualization},''
  \emph{Computer Networks}, vol.~54, no.~5, pp. 862--876, 2010.

\bibitem{Aidi-CNSM2018}
S.~Aidi, M.~F. Zhani, and Y.~Elkhatib, ``On improving service chains
  survivability through efficient backup provisioning,'' in \emph{IEEE/ACM/IFIP
  International Conference on Network and Service Management (CNSM)}, Rome,
  Italy, Nov. 5-9, 2018.

\bibitem{mijumbi2015design}
R.~Mijumbi, J.~Serrat, J.-L. Gorricho, N.~Bouten, F.~De~Turck, and S.~Davy,
  ``Design and evaluation of algorithms for mapping and scheduling of virtual
  network functions,'' in \emph{IEEE Conference on Network Softwarization
  (NetSoft)}, 2015.

\bibitem{tusa2018multi}
F.~Tusa, S.~Clayman, D.~Valocci, and A.~Galis, ``Multi-domain orchestration for
  the deployment and management of services on a slice enabled nfvi,'' in
  \emph{IEEE Conference on Network Function Virtualization and Software Defined
  Networks}, 2018.

\bibitem{galis2018network}
A.~Galis and K.~Makhijani, ``Network slicing landscape: A holistic
  architectural approach, orchestration and management with applicability in
  mobile and fixed networks and clouds,'' in \emph{IEEE Network Softwarization
  (NetSoft)}, 2018.

\bibitem{RabbaniIEICE13}
M.~G. Rabbani, M.~F. Zhani, and R.~Boutaba, ``On achieving high survivability
  in virtualized data centers,'' \emph{IEICE Transactions on Communications},
  vol. E97-B, no.~1, pp. 10--18, Jan 2014.

\bibitem{ZhangVenice2014}
Q.~Zhang, M.~F. Zhani, M.~Jabri, and R.~Boutaba, ``{Venice: Reliable virtual
  data center embedding in clouds},'' in \emph{IEEE International Conference on
  Computer Communications (INFOCOM)}, Toronto, Ontario, Canada, Apr. 27--Mai 2,
  2014.

\bibitem{Ayoubi2016}
S.~Ayoubi, Y.~Chen, and C.~Assi, ``Towards promoting backup-sharing in
  survivable virtual network design,'' \emph{IEEE/ACM Transactions on
  Networking}, vol.~24, no.~5, pp. 3218--3231, 2016.

\bibitem{SVNE2013}
M.~R. Rahman and R.~Boutaba, ``{SVNE}: Survivable virtual network embedding
  algorithms for network virtualization,'' \emph{IEEE Transactions on Network
  and Service Management}, vol.~10, no.~2, pp. 105--118, 2013.

\bibitem{Zhani2015Surv}
\BIBentryALTinterwordspacing
M.~F. Zhani and R.~Boutaba, \emph{Survivability and Fault Tolerance in the
  Cloud}.\hskip 1em plus 0.5em minus 0.4em\relax John Wiley \& Sons, Inc, 2015,
  pp. 295--308. [Online]. Available:
  \url{http://dx.doi.org/10.1002/9781119042655.ch12}
\BIBentrySTDinterwordspacing

\bibitem{Sherry2015}
J.~Sherry, P.~X. Gao, S.~Basu, A.~Panda, A.~Krishnamurthy, C.~Maciocco,
  M.~Manesh, J.~a. Martins, S.~Ratnasamy, L.~Rizzo, and et~al.,
  ``Rollback-recovery for middleboxes,'' in \emph{ACM Conference on Special
  Interest Group on Data Communication (SIGCOMM'15)}, 2015, p. 227–240.

\bibitem{Zhang2016OpenNetVMAP}
W.~Zhang, G.~Liu, W.~Zhang, N.~Shah, P.~Lopreiato, G.~Todeschi, K.~K.
  Ramakrishnan, and T.~Wood, ``{OpenNetVM}: A platform for high performance
  network service chains,'' in \emph{Workshop on Hot topics in Middleboxes and
  Network Function Virtualization}, 2016.

\bibitem{hwang2015netvm}
J.~Hwang, K.~K. Ramakrishnan, and T.~Wood, ``Netvm: high performance and
  flexible networking using virtualization on commodity platforms,'' \emph{IEEE
  Transactions on Network and Service Management}, vol.~12, no.~1, pp. 34--47,
  2015.

\bibitem{liu2017design}
G.~Liu, K.~Ramakrishnan, M.~Schlansker, J.~Tourrilhes, and T.~Wood, ``Design
  challenges for high performance, scalable nfv interconnects,'' in \emph{ACM
  Workshop on Kernel-Bypass Networks}, 2017, pp. 49--54.

\bibitem{ToussainHipNet2018}
A.~Toussain, M.~Hawari, and T.~Clausen, ``{Chasing Linux Jitter Sources for
  Uncompressed Video},'' in \emph{IEEE/IFIP/ACM International Workshop on
  High-Precision Networks Operations and Control (HiPNet)}, 2018.

\bibitem{HongHipNet2018}
J.~Hong, S.~Jeong, J.-H. Yoo, and J.~Hong, ``{Design and Implementation of
  eBPF-based Virtual TAP for Inter-VM Traffic Monitoring},'' in
  \emph{IEEE/IFIP/ACM International Workshop on High-Precision Networks
  Operations and Control (HiPNet)}, 2018.

\bibitem{Kumar2019}
P.~Kumar, N.~Dukkipati, N.~Lewis, Y.~Cui, Y.~Wang, C.~Li, V.~Valancius,
  J.~Adriaens, S.~Gribble, N.~Foster, and et~al., ``Picnic: Predictable
  virtualized nic,'' in \emph{ACM Special Interest Group on Data Communication
  (SIGCOMM'19)}, 2019, p. 351–366.

\bibitem{Chowdhury2014Payless}
S.~R. Chowdhury, M.~F. Bari, R.~Ahmed, and R.~Boutaba, ``Payless: A low cost
  network monitoring framework for software defined networks,'' in \emph{IEEE
  Network Operations and Management Symposium (NOMS)}, May 2014, pp. 1--9.

\bibitem{Shu2016}
Z.~Shu, J.~Wan, J.~Lin, S.~Wang, D.~Li, S.~Rho, and C.~Yang, ``Traffic
  engineering in software-defined networking: Measurement and management,''
  \emph{IEEE Access}, vol.~4, pp. 3246--3256, 2016.

\bibitem{kim2015inband}
C.~Kim, A.~Sivaraman, N.~Katta, A.~Bas, A.~Dixit, and L.~J. Wobker, ``In-band
  network telemetry via programmable dataplanes,'' in \emph{ACM SIGCOMM}, 2015.

\bibitem{InBandTel2016}
C.~Kim, P.~Bhide, E.~Doe, H.~Holbrook, A.~Ghanwani, D.~Daly, M.~Hira, and
  B.~Davie, ``{In‐band Network Telemetry (INT)},'' Jun. 2016.

\bibitem{Haoyu2013}
H.~Song, ``Protocol-oblivious forwarding: Unleash the power of sdn through a
  future-proof forwarding plane,'' in \emph{ACM SIGCOMM Workshop on Hot Topics
  in Software Defined Networking (HotSDN'13)}, 2013, p. 127–132.

\bibitem{Li2017}
S.~{Li}, D.~{Hu}, W.~{Fang}, S.~{Ma}, C.~{Chen}, H.~{Huang}, and Z.~{Zhu},
  ``{Protocol Oblivious Forwarding (POF): Software-Defined Networking with
  Enhanced Programmability},'' \emph{IEEE Network}, vol.~31, no.~2, pp. 58--66,
  2017.

\bibitem{ElkhatibIntentCNSM2017}
Y.~Elkhatib, G.~Coulson, and G.~Tyson, ``Charting an intent driven network,''
  in \emph{International Conference on Network and Service Management (CNSM)},
  Nov 2017.

\bibitem{clemm-nmrg-dist-intent-01}
\BIBentryALTinterwordspacing
A.~Clemm, L.~Ciavaglia, and L.~Z. Granville, ``{Clarifying the Concepts of
  Intent and Policy},'' Internet Engineering Task Force, Internet-Draft
  draft-clemm-nmrg-dist-intent-01, Jul. 2018, work in Progress. [Online].
  Available:
  \url{https://datatracker.ietf.org/doc/html/draft-clemm-nmrg-dist-intent-01}
\BIBentrySTDinterwordspacing

\bibitem{Gessner2012}
D.~Gessner, A.~Olivereau, A.~S. Segura, and A.~Serbanati, ``Trustworthy
  infrastructure services for a secure and privacy-respecting internet of
  things,'' in \emph{IEEE International Conference on Trust, Security and
  Privacy in Computing and Communications}, June 2012, pp. 998--1003.

\bibitem{Takabi2010}
\BIBentryALTinterwordspacing
H.~Takabi, J.~B. Joshi, and G.~Ahn, ``Security and privacy challenges in cloud
  computing environments,'' \emph{IEEE Security and Privacy}, vol.~8, pp.
  24--31, 11 2010. [Online]. Available:
  \url{doi.ieeecomputersociety.org/10.1109/MSP.2010.186}
\BIBentrySTDinterwordspacing

\bibitem{sun2011surveying}
D.~Sun, G.~Chang, L.~Sun, and X.~Wang, ``Surveying and analyzing security,
  privacy and trust issues in cloud computing environments,'' \emph{Procedia
  Engineering}, vol.~15, pp. 2852--2856, 2011.

\end{thebibliography}

\textbf{Mohamed Faten Zhani} is Associate Professor of Software and IT Engineering at ÉTS Montreal (Canada).
His research interests include cloud computing, network function virtualization, software-defined
networking and resource management. He is co-editor of the IEEE Communications Magazine feature
series on Network Softwarization and vice-chair of the IEEE Network Intelligence Initiative.\\

\textbf{Hesham ElBakoury} is a 35-year veteran in the telecommunications and data networking industry with an extensive background and expertise in the architecture, design and development of Distributed Systems and Broadband Access, Enterprise and Telco Communications Systems. He was a Principal Architect in Futurewei in the network research Lab at the time of writing of this paper.

\end{document}